\documentclass[10pt,onecolumn]{IEEEtran}
\IEEEoverridecommandlockouts

\usepackage{amsmath,amsthm}
\usepackage{newtxmath}   

\usepackage{graphicx}
\usepackage{float}
\usepackage{stfloats}
\usepackage{url}
\usepackage{anyfontsize}
\usepackage{enumitem}
\usepackage{bm}
\usepackage[font=footnotesize]{caption}
\usepackage{hyperref}   


\theoremstyle{plain}

\theoremstyle{definition}

\newcommand{\R}{\mathbb{R}}

\newcommand{\ub}{\bm{u}}

\title{\LARGE \bf
Optimal Control and Closed-Loop Stability of Droplet Transport in a Microchannel 
}
\author{Rajneesh Anand\textsuperscript{1} and Mayuresh V. Kothare\textsuperscript{1*}\thanks{*Corresponding Author: mvk2@lehigh.edu}\\[4pt]
\normalsize\textsuperscript{1}Department of Chemical and Biomolecular Engineering, Lehigh University, Bethlehem, PA 18015, USA\\
}

\IEEEoverridecommandlockouts
\begin{document}

\makeatletter
\let\ref\@refstar
\makeatother

\maketitle
\thispagestyle{empty}
\pagestyle{empty}

\begin{abstract}

Understanding the efficient transport of fluid droplets in confined geometries has been a domain of interest for industrial applications in recent times. In the present work, we focus on designing control strategies that optimally steer droplet motion. Here, we apply optimal control framework to the  droplet transport problem in a microchannel based on lubrication theory that minimizes viscous dissipation. Two complementary modeling routes are adopted: a reduced-order ordinary differential equation (ODE) model optimized via Pontryagin's Maximum Principle, and a full nonlinear partial differential equation (PDE) model optimized using a Covariance Matrix Adaptation-Evolutionary Strategy. By parameterizing target displacement, droplet size, and capillary number, we uncover two distinct optimal transport regimes: a “translate–relax” strategy for short distances and a “compact–translate–relax” strategy for longer targets. In continuum mechanics, the competition between surface forces and cumulative viscous dissipation decides the optimal transport strategies. We further show that the displacement range over which the reduced-order controller transfers to the continuum model is governed by capillary stiffness.
We finally address closed-loop stability by adapting a control Lyapunov function (CLF) framework to the reduced-order dynamics. We demonstrate that the terminal cost which penalizes deviation from the target, serves as a cost-compatible CLF on the physical domain, and prove that the CLF-compatible feedback exponentially stabilizes the target state.

\end{abstract}
\noindent

\section{INTRODUCTION}

In recent years, transport and dynamics of fluid interfaces in single phase and multiphase have been areas of major research interest in many engineered systems, ranging from biological processes \cite{ref1}, microfluidics \cite{ref2}, and interfacial physics \cite{anand2023} to turbulent regime \cite{ref3, anand} and confined bluff-body flows \cite{AnandTandem}. When such systems operate far from equilibrium, their dynamics become highly nonlinear and sensitive to spatiotemporal controls \cite{ref1}. This motivates the use of optimal control as an interpretable framework to design spatiotemporal actuation strategies that balance transport efficiency, deformation, and cost. A broad class of navigation problems involving guiding aircraft or drones has been formulated to minimize transport time or energy \cite{ref4}. Recently, a similar framework has been extended to micro-swimmers and active particles. Owing to their small size and driving force, these systems are influenced by the surrounding flow and strategically use it to reach target \cite{ref5}. On the multicellular scale, patterned substrates and external fields have enabled controlled migration of monolayers which offers new possibilities for the control of active biological assemblies \cite{ref6, Endresen2021}.

Classical optimal transport theory, originating from Monge \cite{ref7} proposed a highly nonlinear-nonconvex problem of moving mass from pile of earth to fill the holes of earth, later, Kantorovich relaxed the problem by adding local mass conservation and provided the direction to move mass or fluid distributions between given states while minimizing energetic cost over past centuries \cite{Kantorovich}. Later, Benamou-Brenier shapes the problem into a computational fluid dynamics and provided a different domain of interest where optimal control can be utilized for dynamical systems \cite{Benamou}. Still there was a lack of proper theoretical study which unifies geometrical aspects with complex dynamical system, Villani has huge contribution to shape the optimal control protocols in term of theoretical contributions \cite{ref8}. However, these studies focus on isolated agents or multi-agents \cite{multiagent, optimalparticle} settings, and several other studies, such as, how to optimally guide extended, deformable materials that evolve through complex fluid domain \cite{ref9, Alvarado2026}, thermal transient dynamics \cite{bleris2005reduced}, and model predictive control in membrane microreactors \cite{Kothare2}. In Ref. \cite{ref9,ref10}, it has been studied for active drops (e.g., bacteria) which have an internal activity to drive the drops. 
A recent review \cite{Alvarado2026} confirms that optimal control in soft and active matter has been predominantly applied to systems with internal driving mechanisms; active Brownian particles steered by light \cite{colabrese2017, muinos2021}, active nematics controlled via spatiotemporal activity fields \cite{norton2020optimal, ghosh2025}, and contractile gels modulated by optogenetic signals \cite{clarke2025, liu2025optogenetic}. All of which rely on the system's intrinsic activity rather than an externally imposed pressure or flow field.

For a physical fluidic system where the control input is a  pressure gradient or an externally imposed gradient rather than intrinsic activity, this class of problem remains unexplored.
This gap leads to an unresolved question: how can we design and explore optimal strategies to efficiently transport fluids (droplets) under external fields in a microchannel? Here, we address this problem in the setting of translating a droplet through controlled spatiotemporal driving force. The droplet movement is governed by lubrication theory and capillary effects. By utilizing these constraints into the formulation and defining transport cost in terms of energetic dissipation and deformation penalties, we pose the question: how to optimally move droplet  in a microchannel.

However, finding the optimal control for droplet movement is not enough, one must also ensure that the controller is stable, since a finite-horizon optimal strategy can fail to reach the target under perturbations. Stability of finite-horizon optimal control is not automatically guaranteed; classical counterexamples show that optimality over a finite horizon can produce unstable closed-loop behavior~\cite{kalman1960contributions}. The resolution, developed by Jadbabaie~\cite{jadbabaie2001thesis}, employs a control Lyapunov function (CLF) as the terminal cost to ensure closed-loop stability without explicit terminal constraints. This framework has been widely adopted in nonlinear systems and model predictive control~\cite{bleris2005reduced,mayne2000constrained}, and extended to safety-critical settings~\cite{ames2019control}. In microfluidic systems, feedback control of droplets has been explored via PI control~\cite{Naz2025}, AI-assisted methods~\cite{Guo2024}, and model predictive control (MPC) of slug flow~\cite{ramirez2024dynamic}, but these approaches lack optimal control and formal stability analysis. Optimal control of active nematics~\cite{ref10, norton2020optimal,ghosh2025} focuses on computing open-loop controls without stability analysis. Despite growing interest, CLF-based stability analysis for optimal microfluidic controllers has not been addressed.

To make this problem tractable and interpretable, we project the continuum droplet dynamics onto a lower order of modes, leading to a low-dimensional system of ordinary differential equations (ODEs) describing two states (position and shape) of the droplet. We have interestingly observed two distinct optimal transport regimes, corresponding to translate--relax (TR) and compact--translate--relax (CTR) strategies. Here, CTR is similar to gather-move-spread strategy for active drops \cite{ref10}.  These strategies arise from the competition between transport efficiency and transport cost. Full numerical simulations of the continuum model (partial differential equations (PDEs)) coupled with a gradient-free evolutionary optimization algorithm, confirming the qualitative predictions of the reduced-order description.
We then provide a complete stability analysis for reduced-order model (ODE) by adapting the CLF framework~\cite{jadbabaie2001thesis}. Specifically, we establish: (i) the terminal cost is a valid CLF on the physical domain; and (ii) the CLF-compatible feedback yields exponential stability to the target. Our analysis exploits the driftless, control-affine structure of the droplet dynamics.
Together, these results provide an interpretable framework for understanding optimal droplet transport and the trade-offs between efficiency, control efforts, and robustness.

\section{MATHEMATICAL MODEL FOR OPTIMAL DROPLET TRANSPORT}
\subsection{Depth-integrated Continuity and Flux}

\begin{figure*}
    \centering
    \includegraphics[width=\linewidth]{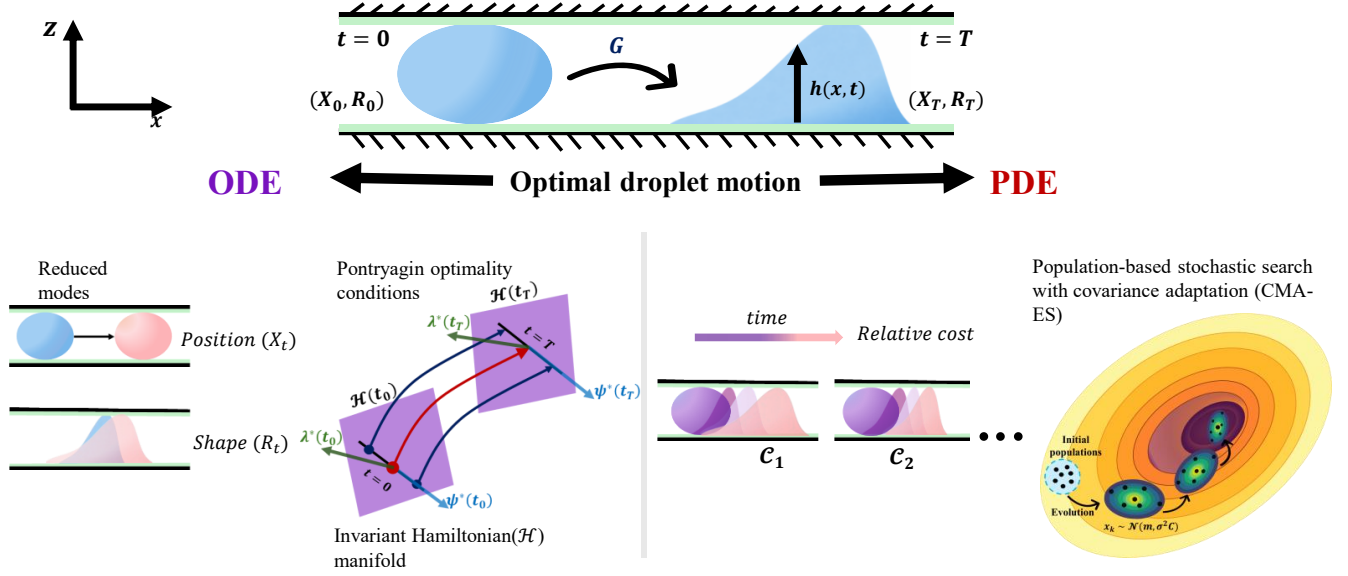}
    \caption{A schematic illustration of optimal motion of a droplet from its initial position and size $(X_0,R_0)$ to a target state $(X_T,R_T )$ while minimizing a cost, such as viscous dissipation in a microchannel. The control $(G)$ shapes $h(x,t)$. We utilize two parallel approaches: (i) a reduced-order ODE model derived from position and shape modes and solved using Pontryagin’s Maximum Principle (PMP), and (ii) a full PDE formulation optimized with a gradient-free evolutionary strategy (CMA-ES) using adaptive Gaussian sampling ($\mathcal{N}(m, \sigma^2C)$, where $m$ = mean, $\sigma^2$ = variance, $\mathcal{C}$ = covariance matrix).}
    \label{fig1}
\end{figure*}

We begin with incompressible flow $\nabla \cdot \text{u} = 0$ \textbf{(SI Appendix)}. In the microchannel geometry, the droplet is described by a height field $h(x,t)$ that measures the local thickness of the droplet phase. We describe a two-dimensional (2D) droplet dynamics moving axially (in the $x$-direction) on a thin film of oil on the solid surface (see Fig. \ref{fig1}), illustrated by lubrication theory \cite{ref9, ref10, ref11, Anand2026surfactant}. 
Here the droplet is separated from the channel walls by a continuous phase (oil), so that the lubricating film regularizes contact-line motion rather than eliminating it; the residual contact-line physics is retained in the PDE model through a precursor film and a disjoining pressure (SI Appendix).
We neglect gravity by considering the drop size to be smaller than the capillary length. Integrating the continuity equation across the film thickness and applying the kinematic condition at the interface gives the depth-averaged continuity equation (Eq.~1) \textbf{(SI Appendix)},
\begin{equation}
\frac{\partial h}{\partial t} + \frac{\partial q}{\partial x} = 0
\tag{1}
\end{equation}

\noindent
$q(x,t) = \int_{0}^{h} u(x,z,t)\,dz$ is the horizontal volumetric flux per unit width. We specify conservation of volume
\begin{equation}
\int h(x,t)\,dx = 1
\tag{2}
\end{equation}

\noindent
In the lubrication limit (for low Reynolds number), the flow is governed by Stokes balance in the horizontal direction (as $H \ll W \ll L$; $H$, $W$, and $L$ represent height, width, and length of the channel, respectively) 
$-\frac{\partial p}{\partial x} + \mu \frac{\partial^{2} u}{\partial z^{2}} = 0.$ 
The pressure consists of capillary curvature and control. We evaluate the velocity,
$u(x,z,t) = \frac{z^{2}-hz}{2\mu}
\left(
\frac{d}{dx}\left(\gamma h_{xx} + Gh\right)
\right),$
where $\mu$ is the viscosity. Integrating it over the film thickness gives the flux (Eq.~3) (Here, $h_x = \partial_x h, h_{xx} = \partial_x^2 h, \And h_{xxx} = \partial_x^3 h$)
\begin{equation}
q(x,t) = -\frac{h^{3}}{12\mu} \left( \gamma h_{xxx} + \partial_x(G(x,t)h)\right)
\tag{3}
\end{equation}

Here, $G(x,t)$ is the control. It is an imposed axial pressure gradient (equivalently, the driving force per unit volume), which may arise from any drive coupling to a property contrast between the droplet and the carrier phase including hydrostatic, centrifugal, magnetic, or dielectrophoretic.
In the current model, Eq.~1 \& Eq.~3 together with boundary and initial condition determines the droplet motion once the control is specified. For finite droplets, additional conditions near the contact line, such as partial slip, prewetting films, or contact angles need to be included \cite{ref12}. In this formulation, surface tension, $\gamma$ sets the curvature of the interface, while the external pressure gradient (control) acts as a spatiotemporal driving term that modifies the effective flux. For convenience, we adopted a simplified form of control input. We describe the control using two unsteady parameters a mean $G_{0}(t)$ and a spatial variation $\Delta G(t)$. This creates a low-dimensional control space that is easy to interpret and suitable for optimization.
We assume the droplet occupies a finite interval on the x-axis, with its left and right edges at $x = a(t)$ and $x = a(t) + R(t)$, respectively. The midpoint of the droplet is therefore
$x_{c}(t) = a(t) + \frac{R(t)}{2}$ (Fig. S1).
We impose linear variation of the control (Eq.~4) to introduce a spatial dependence 
\begin{equation}
G(x,t) = G_{0}(t) + \Delta G(t)\left( \frac{x - x_{c}(t)}{R(t)} \right)
\tag{4}
\end{equation}

However, the droplet motion is passive, and its evolution arises from a competition between externally imposed viscous stresses and restoring forces due to surface tension. The relative strength of these two effects can be measured using a capillary number (Eq.~5). Small values $Ca_{G} \ll 1$ correspond to surface-tension-dominated regime, the droplet remains nearly spherical; $Ca_{G} \gg 1$ shows strong deformation driven by the imposed pressure gradient. For pressure-driven motion, we define an effective capillary number (\textbf{SI Appendix})
\begin{equation}
Ca_{G}
=
\frac{1}{T}
\int_{0}^{T}
\frac{ |\Delta G|\,R(t)^{2}}{\gamma}
\,dt
\tag{5}
\end{equation}

\subsection{Optimal Transport Formulation}

The objective of optimal droplet transport is to identify a transient control input that drives the droplet from its initial state to a target state while minimizing a chosen objective (cost) function. In our formulation, the total cost
$\mathcal{J} = \mathcal{W} + \mathcal{T}$:
$\mathcal{W}$ measures the mechanical effort exerted on the system during the motion (Eq.~6), and a terminal cost $\mathcal{T}$, which penalizes any disagreement between the observed and desired state (Eq.~7). Under the lubrication approximation, the dominant source of dissipation arises from viscous shear, scales as $\mu(\partial_{z}u)^{2}$. The accumulated mechanical work can be expressed as an integral over shear-induced dissipation.
\begin{equation}
\mathcal{W}
=
\int_{0}^{T} dt
\int
\left(
\frac{h^{3}}{12\mu}
\left(
\partial_{x}(Gh + \gamma\,\partial_{xx}h)
\right)^{2}
\right)
dx
\tag{6}
\end{equation}
\begin{equation}\label{eq:terminal_cost}
\mathcal{T}
=
\mathcal{A}
\left(
\frac{X(T) - X_{T}}{X_{T}}
\right)^{2}
+
\mathcal{B}
\left(
\frac{R(T) - R_{T}}{R_{T}}
\right)^{2}
\tag{7}
\end{equation}

The energetic cost associated with the droplet translation is always non-negative, because viscous dissipation cannot generate energy. Moreover, we consider a fixed transport duration $T$, although other formulations such as minimal-time control are possible. To encode the objective of droplet movement to a target distance and achieve a desired final shape, we add a terminal penalty (Eq.~7).
We show that $\mathcal{T}$ satisfies
the conditions of a control Lyapunov function (CLF) for the
reduced-order dynamics, with an explicit exponential
convergence rate \textbf{in section VI}.
$X_{T}$ and $R_{T}$ denote the target center of mass (COM) position and drop size (variance) at the end of the control horizon. $\mathcal{A}$ and $\mathcal{B}$ are terminal cost weighting coefficients. The droplet position is defined through its COM,
$X(t) = \frac{\int x\,h(x,t)\,dx}{\int h(x,t)\,dx},$
using Eq.~2, we have $X(t)=\int x\,h(x,t)\,dx$ and $R(t)$ measures its effective size. These quantities are evaluated at $t=T$ to assess how well control achieves the desired task.

\subsection{Optimal Control Framework}

With the known governing equations and formulated cost functional, we now outline optimal control strategy. A central challenge in this problem is that the dynamics are nonlinear, the cost functional is non-convex, and the control enters multiplicatively through the flux, coupling the state and control in a way that precludes standard linear quadratic methods. We address this through two complementary approaches operating at different levels of model fidelity, each chosen for specific methodological reasons. 

\textit{Reduced order optimal control}: We begin by projecting the continuum model onto low order modes that determine the position, width, and overall shape of the droplet. This projection provides a low-dimensional system of ODEs that approximates the evolution of the droplet along a slowly varying manifold by surface tension ($\gamma$) and viscosity ($\mu$). The model removes nonessential drift terms and ensures that the reduced system captures the balance between deformation and translation induced by the driving force. As ODE model is compact and interpretable, we apply Pontryagin’s maximum principle (PMP) to determine control policies analytically and numerically. These reduced equations reveal distinct optimal strategies, including the TR and CTR modes discussed later.

We apply PMP rather than alternative methods such as
dynamic programming (Hamilton--Jacobi--Bellman) or direct
collocation, for a couple of reasons. First, the ODE system (Eq.~9)
is low-dimensional (2 states, 2 controls) and
control-affine, which places it squarely in the regime where
PMP is most effective \cite{Pontryagin}: the first-order necessary conditions
yield an analytically tractable two-point boundary value
problem (BVP). Second, the stationarity
conditions can be solved
in closed form for the optimal controls \cite{controltheory}.
This analytical elimination of the control variables is not
possible with direct numerical methods, and it provides
physical interpretability: the optimal controls are
proportional to the costates (adjoint variables) weighted by
the droplet size, directly revealing how the system balances
translation and deformation.
In contrast, the Hamilton--Jacobi--Bellman (HJB) approach
would require solving a PDE for the value function over the
entire state space. Direct collocation methods
(e.g., pseudospectral or shooting methods) could solve the
BVP numerically but would not yield the closed-form control
expressions or the conserved Hamiltonian structure that
enable analytical characterization of the transport regimes.

\textit{PDE optimal control}: To validate the reduced-order predictions and capture physical effects (e.g., $\gamma$, interfacial curvature) missing from the ODE model, we solve the optimal transport problem directly using the full nonlinear PDE for $h(x,t)$. Unlike a typical forward simulation, the controlled problem is a non-convex and transient BVP where droplet must satisfy an initial state at $t=0$ and a final state at $t=T$, making adjoint-based optimization stiff and difficult. Gradient-based method risk converging to suboptimal solution depending on initialization. 
we instead use a gradient-free evolutionary optimizer. We employ covariance matrix adaptation evolution strategy (CMA-ES) to explore the space of control profiles $G(x,t)$ \cite{CMAES}.  
CMA-ES is a stochastic, population-based method that
maintains and adapts a multivariate Gaussian distribution
$\mathcal{N}(m, \sigma^2 C)$ over the control
parameter space (Fig. 1). At each generation, it samples candidate
control profiles, evaluates the cost by integrating the PDE
forward, ranks the candidates, and updates the $m$, $\sigma$, and $C$ toward lower-cost regions.
This stochastic search allows us to identify the optimal control profile without requiring derivatives of the cost functional.

CMA-ES was chosen over other derivative-free methods for
the following reasons. Compared to genetic algorithms (GA)
or particle swarm optimization (PSO), CMA-ES adapts its
search distribution shape through covariance learning, which
is particularly advantageous when the cost landscape exhibits
anisotropic sensitivity to different control parameters---as
is the case here, where the cost is more sensitive to
$\Delta G(t)$ (which drives translation) than to $G_0(t)$
(which modulates shape). Compared to Bayesian optimization,
CMA-ES scales more naturally to the $O(10^2)$ parameter
dimensions involved and does not require a surrogate model
whose accuracy would be difficult to guarantee for a
PDE-constrained objective. Furthermore, CMA-ES is
invariant under order-preserving transformations of the cost
function and under rotation/translation of the search space \cite{CMAES},
properties that make it robust to the specific scaling of
our problem.

\section{ODE CONTROL}

For a capillary-driven droplet with uniform internal pressure, the Laplace pressure
$p = -\gamma \kappa$ implies that the interface curvature is constant. This leads to a
parabolic height profile, as detailed in \textbf{SI Appendix}
\begin{equation}
h(x,t)
=
\frac{6}{R(t)^3}
\left[
\frac{R(t)^2}{4} - (x - X(t))^2
\right]
\tag{8}
\end{equation}

This motivates a reduced-order description in which the continuum height field is expressed
in terms of the droplet position $X(t)$ and its effective width $R(t)$. The idea is to
introduce this parametrized profile into the lubrication equation and project the
residual dynamics onto low-order moments of $h(x,t)$, using Galerkin projection.
The first moment gives the evolution of the COM,
$X(t)=\int x\,h(x,t)\,dx$, while the second central moment,
$\Delta(t)=\int (x-X(t))^2 h(x,t)\,dx$, gives a measure of the droplet’s instantaneous
size or deformation. Using the parabolic profile (Eq.~8), $\Delta(t)$ reduces to a function of
$R(t)=\sqrt{20\Delta(t)}$, allowing the PDE to collapse onto a pair of coupled ODEs
for $X(t)$ and $R(t)$. Using Eq.~1 and Eq.~3, the evolution of $X(t)$ and $R(t)$ can be
obtained by $\dot{X}(t)=\int_{X(t)-R(t)/2}^{X(t)+R(t)/2} q(x,t)\,dx$, and
$\dot{\Delta}(t)=2\int_{X(t)-R(t)/2}^{X(t)+R(t)/2} (x-X(t))\,q(x,t)\,dx$. The integration
limits $X(t)\pm R(t)/2$ correspond to the left and right edges of the droplet, where
$h(x,t)=0$. Carrying out above Galerkin projection for pressure-driven system yields the
nonlinearly coupled ODEs for the droplet’s position and size (Eq.~9) \textbf{(SI Appendix),}
\begin{equation}\label{eq:dynamics}
\dot{X}(t)
=
\frac{9\,\Delta G(t)}{70\,\mu\,R(t)^4},
\qquad
\dot{R}(t)
=
-\frac{6\,G_0(t)}{7\,\mu\,R(t)^4}
\tag{9}
\end{equation}

The mean component of the imposed pressure gradient $G_0(t)$ primarily controls the
droplet size, while the gradient component $\Delta G(t)$ determines its translation.
The reduced dynamics (Eq.~9) contain no surface tension and drift terms, which signifies
that no motion occurs unless a control signal is applied. This is advantageous because
it ensures controllability \textbf{(SI Appendix)} \cite{ref15}. 
For ease, we study a fixed-endpoint
problem, where the droplet reaches specified values $X(T)=X_T$ and $R(T)=R_T$ at the
final time. Under this constraint, the cost functional reduces to the accumulated
dissipation,
$\mathcal{W}=\int_0^T \mathcal{L}(t)\,dt$, the dissipation rate (\textbf{see SI Appendix}) is

\begin{equation}
\mathcal{L}(t)
=
\frac{1}{\mu R(t)^6}
\left(
\frac{18}{35}G_0(t)^2 + \frac{27}{154}\Delta G(t)^2
\right)
\tag{10}
\end{equation}

To determine the optimal controls, we apply Pontryagin’s Maximum Principle (PMP).
Introducing co-states $p_X(t)$ and $p_R(t)$ to enforce the reduced dynamics,
the control Hamiltonian
\(\left([\,p_X\;\;p_R\,]
[\dot{X}\;\;\dot{R}\,]^T
- \mathcal{L}\right)\)
takes the form
\begin{equation}
\mathcal{H}
=
p_X \frac{9\,\Delta G}{70\,\mu R^4}
-
p_R \frac{6\,G_0}{7\,\mu R^4}
- 
\mathcal{L}(t)
\tag{11}
\end{equation}

Optimality requires that $\mathcal{H}$ be maximized with respect to the control inputs \textbf{(SI Appendix)}.
Solving $\partial \mathcal{H}/\partial G_0 = 0$ and
$\partial \mathcal{H}/\partial \Delta G = 0$, gives optimal control
$(G_0^*(t),\Delta G^*(t)) =
\left(-\frac{5}{6}p_R R^2,\,\frac{11}{30}p_X R^2\right)$.
Substituting these expressions back into the Hamiltonian gives a conserved quantity,
allowing the state and costate dynamics to be written in canonical form
$(\dot{X},\dot{R})=(\partial \mathcal{H}/\partial p_X,\partial \mathcal{H}/\partial p_R)$
and $(\dot{p}_X,\dot{p}_R)=(-\partial \mathcal{H}/\partial X,-\partial \mathcal{H}/\partial R)$.
Since the system is translationally invariant,
$\partial \mathcal{H}/\partial X = 0$, implying that $p_X$ is conserved along the
optimal trajectory. The resulting ODE--costate system determines the optimal route by
which the droplet changes its size and position over time.

\begin{figure}
    \centering
    \includegraphics[width=0.75\linewidth]{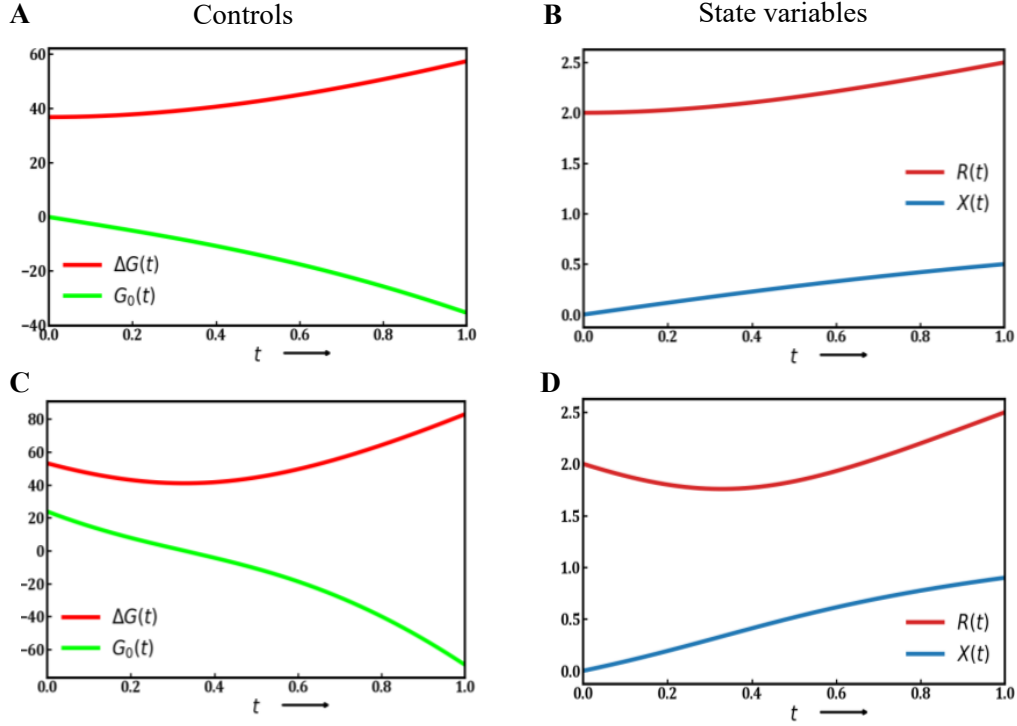}
    \caption{ODE optimal control. (A—B) Optimal time histories of the control inputs depicts the “TR” regime for $X_T=0.5,R_0=2.0, \And R_T=2.5$. $\Delta G(t)$ governs translation while $G_0 (t)$ regulates the droplet size are shown in A. The corresponding evolution of the droplet width $R(t)$ and position $X(t)$ are shown in B. (C—D) Optimal controls and state trajectories represents the “CTR” strategy for $X_T=0.75,R_0=2.0,\And R_T=2.5$. In this regime, the optimal solution shows initial compaction phase, followed by a translation phase, after which the control switches to a relaxation mode to reach the $X_T  \And R_T$, shown in C. The corresponding trajectories show a non-monotonic growth of $R(t)$, as shown in D.}
    \label{fig2}
\end{figure}

The reduced system has been solved analytically to obtain the optimal controls that move the droplet from $(X_0,R_0)$ to the target state $(X_T,R_T)$ within the fixed time interval $T$. These analytical solutions give explicit expressions for $G_{0}(t)$ and $\Delta G(t)$, and therefore for the optimal evolution of $X(t)$ and $R(t)$. Two different cases of such an optimal trajectory have been shown on Fig. \ref{fig2}. For the case $X_{T}=0.5, 0.75$, $R_{0}=2.0$, and $R_{T}=2.5$. Although the first-order optimality conditions allow local minima, our problem has a global minimum since it is convex in the reduced parameter space. This globally minimizing solution is shown in Fig. \ref{fig2}, which represents the least dissipative transport protocol for this task. Despite identical size constraints, the optimal solutions exhibit qualitatively different strategies depending on the required $\frac{X_{T}}{R_{T}}$.

For the smaller target displacement ($X_{T}=0.5$), the optimal solution corresponds to a translate--relax (TR) regime, in which droplet translation and relaxation (spreading) occur concurrently (Fig. \ref{fig2}(A \& B)). In this regime, the droplet moves toward its target position while its size increases monotonically, without any intermediate compaction phase. In a confined microchannel, relaxation increases the droplet’s contact area along the channel, thereby enlarging the region over which shear and pressure-driven flow occur. This leads to enhanced viscous dissipation during translation. However, when the total transport distance is small, the dissipation accumulated over the trajectory remains modest. Under these conditions, actively compacting the droplet would incur an additional energetic cost that cannot be recovered over the short transport path. As a result, the optimal strategy avoids unnecessary shape preparation and directly couples translation and relaxation.

In contrast, for large target displacements ($X_{T}=0.75$), the optimal solution transitions to a compact--translate--relax (CTR) regime characterized by a non-monotonic evolution of droplet size (Fig. \ref{fig2}(C \& D)), a similar flow profile to \cite{ref10}. In this regime, the droplet undergoes an initial contractions phase, followed by long-range translation in a compact configuration, and finally relaxes to its prescribed final size near the target location. This behavior arises from a different energetic balance. 
In lubrication-dominated flows, viscous dissipation scales strongly with the droplet’s spatial extent. Bigger size droplets experience shear and pressure gradients over a larger region, leading to greater cumulative dissipation during translation. When the transport distance is large, this dissipation accumulates over time and becomes the dominant contribution to the cost functional. Under these conditions, an upfront energetic investment to compact the droplet is repaid many times over through reduced dissipation during the extended translation phase. The emergence of the TR and CTR regimes demonstrates the dual role of droplet size in confined microchannels. While the initial and final sizes are fixed by boundary constraints, the droplet’s intermediate size evolution remains a free degree of freedom that strongly influences transport efficiency. The optimal strategy depends not on the absolute droplet size, but on the relative importance of transport-induced dissipation, which increases with the required displacement. These results reveal a general physical principle for soft matter transport in viscous confinement: short-range transport favors direct motion without shape preparation, while long-range transport benefits from an initial compaction that minimizes dissipation during translation. The transition between TR and CTR regimes reflects a shift from shape-passive to shape-active transport, driven by the cumulative nature of viscous losses.
While the ODE model provides intuitive physical insights into the droplet dynamics, shape stability requires explicit capillary effects ($\gamma$). We therefore extend the description to the full PDE model.

\section{PDE CONTROL}

The full PDE model allows for deformation of the droplet as it includes curvature changes and shape asymmetry, which cannot be captured by the ODE system. In the PDE model, Eq.~1 and Eq.~3 naturally capture curvature effects, thinning near the contact line, and any higher-order shape variations that occur during the motion. To compute smooth and physically practical control fields, we modify the cost functional that includes three contributions $\mathcal{J} = \mathcal{W} + \mathcal{T} + \mathcal{R}$ \textbf{(SI Appendix)}. The dissipation cost is slightly modified (Eq.~12) in PDE model 

\begin{equation}
\mathcal{W}
=
\int_{0}^{T} dt
\int dx\;
\frac{h^{3}}{12\mu}
\left[
\partial_{x}(Gh + \gamma \partial_{xx}h - \Pi
\right)]^{2}
\tag{12}
\end{equation}

\begin{equation}
\mathcal{R}
=
\frac{\alpha}{N_{c}}
\int_{0}^{T} dt
\left[
(\partial_{t}G_{0})^{2}
+
(\partial_{t}G_{1})^{2}
\right]
\tag{13}
\end{equation}

$\Pi$ is disjoining pressure arises from interactions between the liquid and the solid surfaces, preventing the film from thinning further. Terminal cost remains same as Eq.~7. Brockett’s minimum attention \cite{ref17} regularization cost (Eq.~13) is added to suppress temporal oscillations in the controls that are robust to discretization noise and small perturbations. $\alpha$ is regularization coefficient which represents control actuation time steps and $N_{c}$ depicts 100-equi-spaced (for discretized space) time points which are linearly interpolated. $G1(t)=\Delta G(t)/R(t)$ is the control gradient in absolute coordinates (\textbf{SI Appendix}).
The PDE is discretized and solved using the finite-element method (FEM) using Legacy FEniCS \cite{Fenics}, which provides stable handling of 4th-order nonlinear film dynamics \textbf{(SI Appendix)}. For a candidate control $G(x,t)$, we integrate the PDE from the initial condition $X(0)=0$, $R(0)=R_{0}=2.0$, with viscosity $\mu = 0.5$ and fixed time horizon $T = 1$. Computations are performed on a 1D domain of length $L = 8$, discretized with a uniform spatial resolution $\Delta x \approx 0.01$ ($N = 800$ grid points). Time integration is carried out using a mixed semi-implicit scheme; the nonlinear term is treated explicitly with a second-order Adams–Bashforth method, while the linear stress is handled using Crank–Nicolson. A fixed time step $\Delta t = 0.005$ is used in all simulations (see \textit{Code Availability} section for details).

To determine optimal controls, we use a gradient-free evolutionary algorithm (CMA-ES) \cite{CMAES} to optimize control profiles. CMA-ES iteratively updates control and evaluates cost by solving the PDE. It learns the shape of the objective function by updating a covariance matrix and updates its matrix for better solutions. Initial controls are randomly sampled from a bounded interval to ensure numerical stability, and the optimization is initialized with a moderate population size of 20. The optimization termination criteria are set at reaching a maximum iteration $8 \times 10^4$ and spread in the cost function across the population falls below tolerance, shown in Eq.~14.

\begin{figure*}[t]
    \centering
    \includegraphics[width=\linewidth]{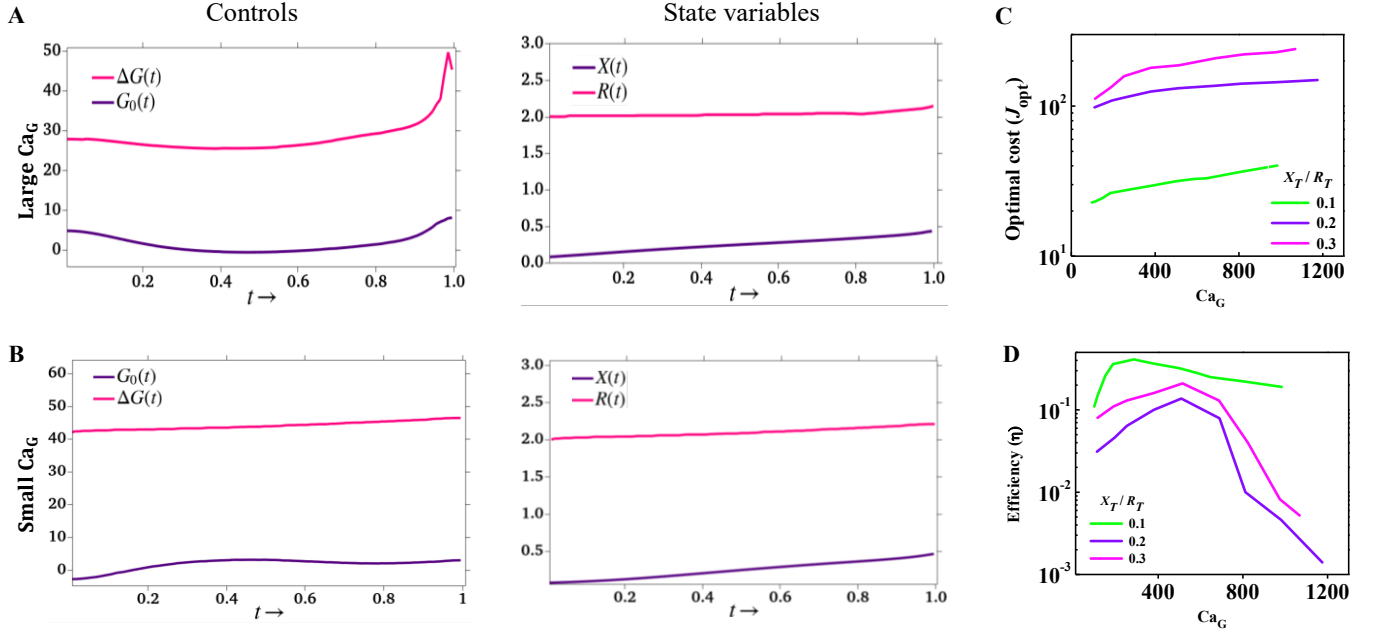}
    \caption{PDE optimal control. (A) Optimal control profiles $G_0 (t)  \And \Delta G(t)$ (Left), and the corresponding trajectories of the droplet $X(t) \And R(t)$ (Right) obtained from full PDE optimization for $\gamma=0.1, Ca_G=1175$. (B) Analogous plot with optimal control profiles $G_0 (t)  \And \Delta G(t)$ (Left), and the corresponding trajectories of the droplet $X(t) \And R(t)$ (Right) has been shown for $\gamma=1.0, Ca_G = 194$. Both A and B are computed by using $X_T=0.5,R_0=2.0, \And R_T=2.5$ kept fixed. The total optimal cost ($\mathcal{J}_{opt}$) (C) and transport efficiency ($\eta$)  (D) are numerically obtained from PDE optimal solution and plotted against $Ca_G$, for varied increasing transport task $(X_T/R_T)$, as shown in different colors (green, purple, magenta). The plots indicate an optimal compromise between viscous forcing and capillary effect, achieved at intermediate capillary numbers.}
    \label{fig3}
\end{figure*}

\begin{equation}
\Delta \mathcal{J} \le \epsilon_{\text{tol}}
\left(
\mathcal{J}_{\text{median}}^{0}
-
\mathcal{J}_{\text{median}}^{\min}
\right)
\tag{14}
\end{equation}
\noindent
$\Delta \mathcal{J} = \max(\mathcal{J}) - \min(\mathcal{J})$ is the current population spread,
$\mathcal{J}_{\text{median}}^{0}$ is the initial median cost, and
$\mathcal{J}_{\text{median}}^{\min}$ is the smallest median cost recorded during the optimization.
We use a tolerance of $\epsilon_{\text{tol}} = 10^{-4}$. It ensures that the optimizer stops only when the internal diversity of the candidate solutions (the search width) becomes negligible compared to the total progress achieved. 

For small surface tension ($\gamma = 0.1$) corresponding to a large capillary number ($Ca_{G}$), the droplet dynamics are dominated by viscous dissipation within the bulk of the fluid. A representative trajectory of the droplet shape and the corresponding control inputs is shown in Fig. \ref{fig3}A, for a target displacement $X_{T} = 0.5$ and $Ca_{G} \approx 1175$. In this regime, the droplet develops an advancing peak and a thin receding tail. 
The optimal control inputs vary smoothly in time and successfully transport the droplet to the target location. Throughout the process, the droplet undergoes smooth and monotonic shape changes, and the resulting control signals are consistent with predictions from the reduced-order ODE model (Fig. \ref{fig2}(A\&B)). However, when surface tension is reduced, the droplet moves smoothly because it can deform easily. If surface tension is too small, the droplet deforms too much, creating fast internal fluid motion and strong shearing. This causes a large amount of energy to be lost through viscosity; the transport efficiency reduces for larger $X(T)$ and total energy cost becomes high.

When the surface tension ($\gamma=1.0$) increases (smaller $Ca_{G}$), the droplet dynamics change qualitatively. As shown in Fig. \ref{fig3}B for $X_{T}=0.5$ and $Ca_{G}\approx 194$, the droplet advances with higher cost but tends to target displacement over the same time interval. In this regime, capillary forces strongly resist shape deformation, leading to a resistance for droplet deformation and comparatively poor impact on transport efficiency than optimal efficiency. 
The observed dynamics are qualitatively different from the waiting-time solutions reported for nonlinear diffusion equations \cite{waiting-time}, as the droplet does not exhibit small-scale oscillations in shape or size. This difference arises because the lubricating film suppresses stick–slip arrest, allowing slow but continuous motion rather than arrest or oscillatory behavior. Since the droplet remains separated from the walls by a thin lubricating film in the microchannel, which suppresses contact-line pinning and stick–slip behavior. As a result, even a small difference between the front and rear shapes of the droplet creates a pressure imbalance along the channel that continuously drives the droplet forward. Transport is therefore governed mainly by pressure-driven flow arising from shape asymmetry rather than by intermittent contact-line motion. Nevertheless, the increased resistance because of higher surface tension leads to viscous dissipation cost and poor performance in the transport task in terms of terminal cost. 
Moreover, in all cases, only the $X(T)$ and $R(T)$ are constrained, while the final droplet shape is not fixed. Once the control input is turned off, surface tension causes the droplet to relax passively back toward a smooth, parabolic shape without extra energy input. Although this relaxation can induce a small residual translation due to shape asymmetry, extending the control protocol would allow this recoil to be compensated at the expense of extra energy.

\begin{figure}
    \centering
    \includegraphics[width=\linewidth]{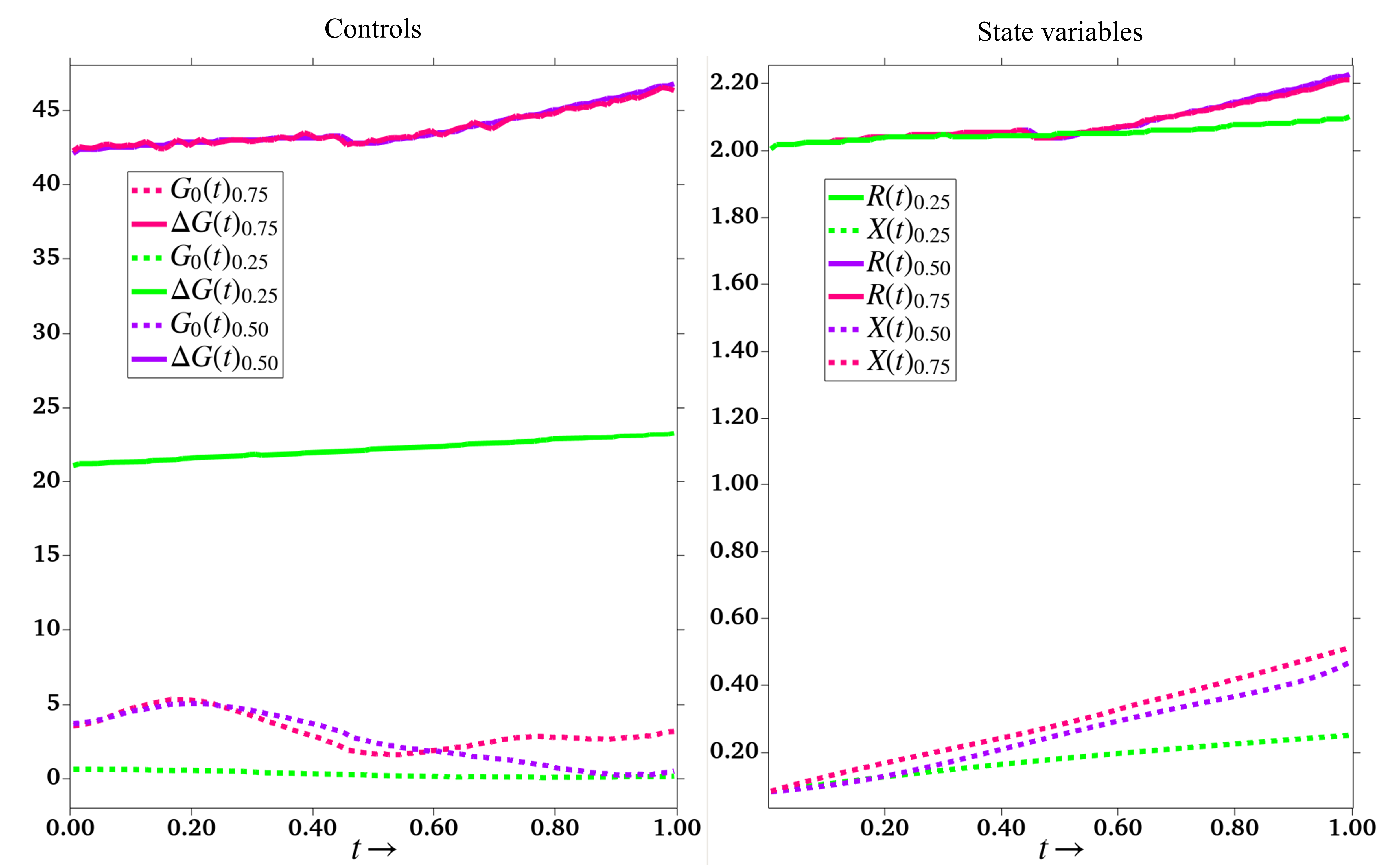}
    \caption{PDE optimal control. Optimal time histories of the control inputs and state variables for the different set of parameters, $X_T=0.25, 0.5, 0.75, \text{ with } R_0 = 2.0 \text{ and } R_T=2.5$. $G_0 (t), \Delta G(t)$ and final droplet size have been recorded and analyzed at different set of $X_T$.}
    \label{fig4}
\end{figure}

The varied optimal strategies obtained upon tuning surface tension and target position suggest a balance between control and capillary forces. 
In contrast to the decreasing cost trends reported in Ref.\cite{ref10}, we observe an increment in optimal cost with capillary number for target displacements considered (Fig. 3C). This apparent discrepancy reflects a difference in the dominant energetic balance. In the present confined geometry, increasing capillary number weakens interfacial stiffness and promotes droplet relaxations, which increases the axial extent of the droplet and amplifying viscous loss and pressure-driven flow. As a result, viscous dissipation accumulated during transport increases with capillary number, leading to a monotonic higher total cost with capillary number for each target displacement.
In the current study, droplet transport is successfully achieved for $(X_{T}=0.25\ \&\ 0.5)$ across all capillary number, $Ca_{G}$ considered.
For a larger target displacement, $X_T = 0.75$, the droplet moves steadily toward the target but does not reach the target and saturating around $X\approx0.53$, shown in Fig.~\ref{fig4}.
As the droplet travels farther, viscous energy losses accumulate along the transport path, gradually reducing the effective controls produced by droplet shape asymmetry. As a result, although the droplet continues to move forward, the remaining propulsion is no longer sufficient to overcome ongoing viscous resistance within the imposed control and time constraints. This lag reflects an energetic limitation on long-distance droplet transport in confined microchannels. 
The optimal cost is the combined effects of droplet translation over the $X_{T}$ and viscous dissipation during transport. As the $X_{T}$ increases, viscous dissipation plays a more important role because the total cost depends on the entire transport history. Since viscous dissipation integrates over time, even a small increment in viscous loss at higher $Ca_{G}$ can lead to a much larger total cost when the droplet travels a long distance. As a result, moderately lower $Ca_{G}$ flow regime becomes the optimal choice for minimizing the overall transport cost. The total cost of the optimal solution is plotted in Fig. \ref{fig3}C.

We also measure efficiency \textbf{(SI Appendix)} which shows non-monotonic dependence on $\gamma$ (Fig. \ref{fig3}D), showing a comparison between droplet deformability and viscous dissipation. 
The observed trends in transport efficiency can be directly interpreted in terms of the TR and CTR transport strategies identified earlier. When the target displacement is small $(X_T=0.25)$, the TR strategy leads to higher transport efficiency. For higher capillary number $(\gamma=0.1)$, the droplet is easily deformable and quickly develops asymmetry. Since the required travel distance is short, the droplet reaches the target before viscous dissipation has time to accumulate significantly. In contrast, at lower capillary number $(\gamma=1.0)$, the droplet is stiffer and resists deformation, so a larger fraction of the control energy is spent overcoming capillary resistance rather than producing translation. As a result, transport efficiency is lower. Overall, for short transport distances, the TR strategy combined with lower surface tension provides a more efficient conversion of input energy into forward motion, leading to an approximately $67\%$ higher efficiency.

When the target displacement is larger $(X_T=0.5)$, the transport efficiency decreases sharply for all surface tensions as viscous losses accumulate over longer transport paths, but the relative performance of different strategies changes. This efficiency penalty at $X_T = 0.5$ is what makes the CTR strategy energetically favorable at larger targets ($X_T = 0.75$).
At high capillary number $(\gamma=0.1)$, the droplet undergoes excessive deformation during motion, generating strong internal flows and sustained viscous dissipation. Consequently, efficiency becomes extremely low. At lower capillary number $(\gamma=1.0)$, viscous dissipation accumulates over time, and it becomes the dominant cost mechanism for long paths. The CTR strategy improves efficiency by minimizing the duration of highly deformed motion, allowing most of the transport to occur in a compact, low-dissipation state. The CTR strategy reduces the time spent in highly dissipative states, allowing translation to occur in a lower-dissipation configuration and increasing efficiency by more than three orders of magnitude compared to the high capillary number case. However, at very high surface tension (too small capillary number), the droplet is too stiff to deform efficiently and leads to comparatively lower efficiency than optimal. Consequently, maximum efficiency is achieved at intermediate surface tension, where sufficient deformation enables effective pressure-driven translation while excessive viscous dissipation is suppressed.

\section{ODE-Derived Controllers on the PDE System}\label{crossvalidation}

To assess whether the analytically derived ODE controls can guide the full continuum (PDE) dynamics, we apply the PMP-derived optimal policy in a receding-horizon fashion to the PDE system. At each time step, the instantaneous PDE state $(X(t), R(t))$ is measured, the ODE costate equations are solved forward over the remaining horizon, and the resulting optimal controls $(G_0^* , \Delta G^*)$ are applied to the PDE for one step before re-planning. This feedback tests the fidelity of the two-mode Galerkin reduction under realistic conditions where surface tension, interfacial curvature, and higher-order deformation modes are all present. Results for three target displacements are shown in Fig. 5. For $X_T = 0.25$, the droplet reaches $95\%$ of the target displacement, confirming that the reduced-order controller captures the dominant transport physics in the translate–relax regime. For $X_T = 0.50$, the droplet attains $\sim 90\%$ of the target before encountering numerical instability near the terminal time. For $X_T = 0.75$, the PDE diverges at $t \approx 0.54$ (mid-horizon) as height goes negative, reaching only $59\%$ of the target.
Notably, this limitation is not specific to only the ODE-derived controller. The full PDE optimization via CMA-ES, operating on the same discretization with unrestricted control freedom (at $\gamma =0.1$), also stagnates near $X \approx 0.53$ for $X_T = 0.75$ (Fig. 4), confirming that the maximum achievable displacement is limited by the cumulative viscous dissipation. The limitation shows up numerically as a loss of film positivity near the contact line, since sustaining translation at larger $X_T$ requires forcing that steepens the interface faster than capillary relaxation can restore it.

\begin{figure*}[t]
    \centering
    \includegraphics[width=\linewidth]{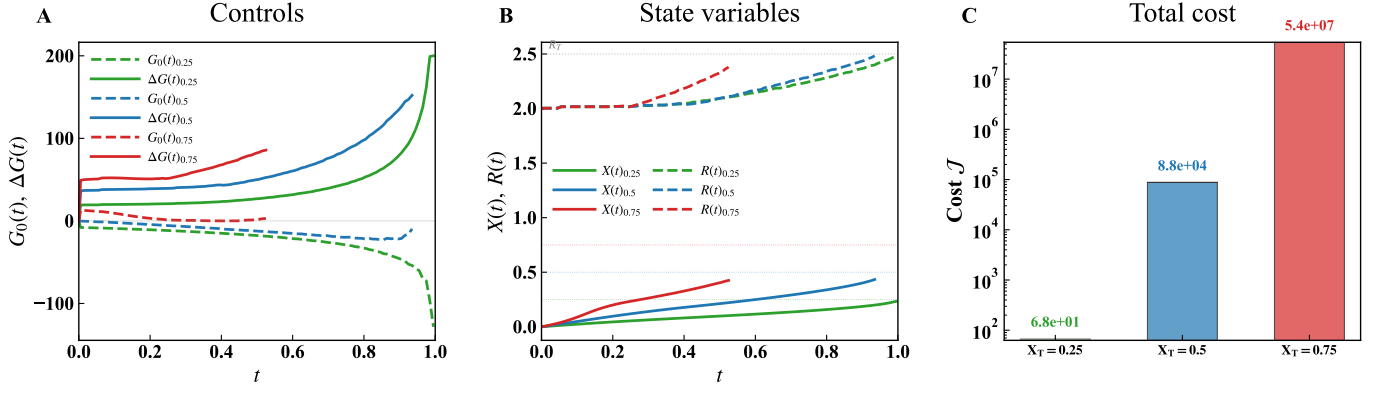}
    \caption{Cross-validation of the ODE-derived receding-horizon PMP controller on the full PDE system for $X_T = 0.25, 0.50, \text{and } 0.75$, with $R_0 = 2.0$, $R_T = 2.5$, $\gamma = 0.1$, and $\mu = 0.5$. (A) Control profiles $G_0(t)$ and $\Delta G(t)$ computed from the PMP optimality conditions using the instantaneous PDE state. (B) State trajectories $X(t)$ and $R(t)$. Horizontal dashed lines indicate the respective target displacements. For $X_T = 0.75$, the simulation terminates at $t \approx 0.54$. (C) Total cost $\mathcal{J}$ on a logarithmic scale, increasing by roughly three orders of magnitude per step from $X_T = 0.25$ ($\mathcal{J} \approx 6.8 \times  10^1$) to $X_T = 0.75$ $(\mathcal{J} \approx 5.4 \times 10^7)$.}
    \label{fig5}
\end{figure*}

This progressive degradation is physically consistent with the energetic saturation observed in the full PDE optimization (Fig. 4), where even CMA-ES with direct PDE evaluation cannot drive the droplet beyond $X \approx 0.53$ for $X_T = 0.75$. In both cases, the underlying cause is the same; as the target displacement increases, the controller demands stronger forcing, which excites higher-order deformation modes i.e., front–back asymmetry, interfacial thinning near the contact line, that the two-mode Galerkin projection does not represent. The comparison confirms that the ODE-derived transport strategies transfer meaningfully to the continuum model for moderate displacement ratios $(X_T/R_T \leq 0.2)$, and simultaneously identifies a quantitative fidelity boundary. The reduced-order controller can no longer account for the shape complexity that the full PDE develops, beyond $(X_T/R_T \approx 0.2)$.

\begin{figure}[t]
\centering
\includegraphics[width=\linewidth]{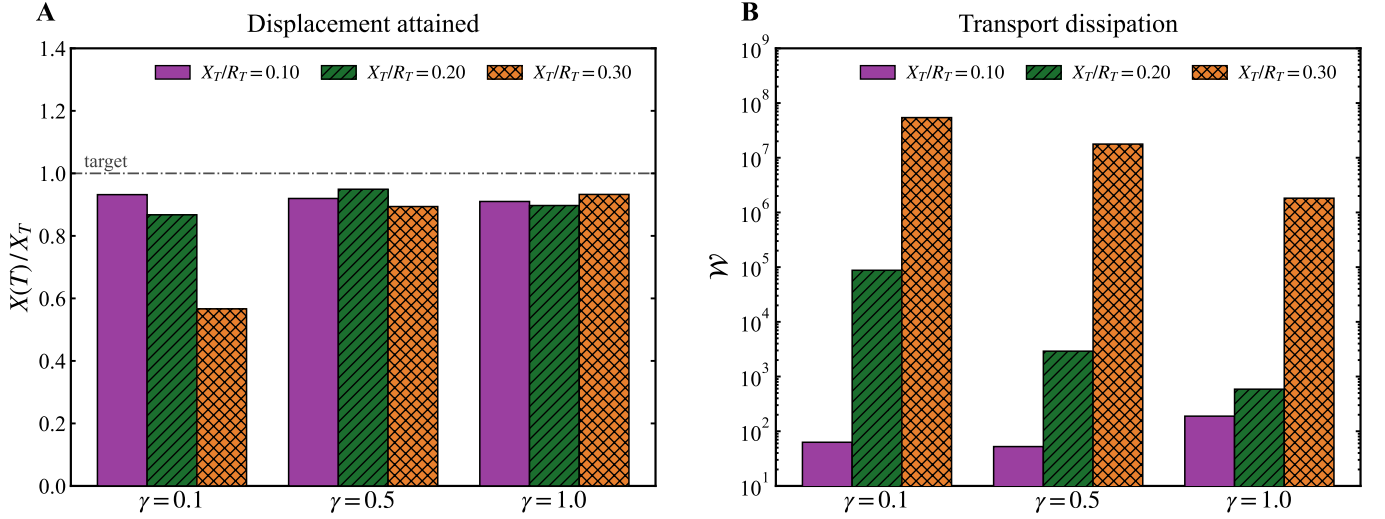}
\caption{Effect of capillary stiffness on reduced-order controller
transfer to a full nonlinear system (PDE). (A) Fraction of the
target displacement attained and (B) accumulated dissipation
$\mathcal{W}$, for three surface tensions $\gamma \in (0.1, 0.5, 1.0)$ and three transport ratios
$X_T/R_T \in (0.1, 0.2, 0.3)$, with
$R_0 = 2.0$, $R_T = 2.5$, and $\mu = 0.5$.}
\label{fig:gamma_sweep}
\end{figure}

The reduced-order controller is extended to the PDE system for varying surface tension $\gamma = 0.1$, $0.5$, and $1.0$ (Fig.~\ref{fig:gamma_sweep}). The fidelity boundary identified above is specific to the weakly capillary regime. At $\gamma = 0.1$ the attained displacement collapses from $95\%$ to $59\%$ as $X_T/R_T$ increases from $0.10$ to $0.30$, whereas at $\gamma = 0.5$ and $1.0$ it remains near $90\%$ across the full range. For $X_T/R_T = 0.30$, increasing $\gamma$ from $0.1$ to $1.0$ delays
the onset of film rupture from $t \approx 0.54$ to $t \approx 0.97$ while reducing the accumulated dissipation. The transport efficiency $X(T)/\mathcal{W}$ is maximized at intermediate $\gamma$ for short displacements and at high $\gamma$ for long ones, consistent with the non-monotonic dependence on capillary number reported in Fig.~\ref{fig3}. Capillary restoring forces therefore
extend the range over which the reduced-order controller transfers by suppressing the interfacial deformation that the two-mode projection does not represent. Detailed plots are shown in the \textbf{SI Appendix, Fig.~S3}.

\section{Closed-Loop Stability of Optimal Transport}\label{sec:stability}

The optimal controls derived in the preceding sections
minimize viscous dissipation over a fixed time horizon, but
optimality alone does not guarantee that the droplet will
converge to the target under perturbations. Previous studies
show that finite-horizon optimal controllers can produce
unstable closed-loop behavior~\cite{kalman1960contributions}. This
distinction is critical for microfluidic applications, where
noise in pressure actuation or imprecise initial conditions are
unavoidable. We therefore ask: are the TR and CTR strategies
inherently stabilizing? To answer this, we adapt the control
Lyapunov function (CLF) framework~\cite{jadbabaie2001thesis} to
the reduced-order dynamics~(9).
The PMP-derived controls (Section III) solve a finite-horizon problem with fixed endpoint constraints, yielding the minimum-dissipation trajectory. We show that the reduced-order ODE system admits a state-feedback law on the physical domain $\mathcal{D}$ under which any prescribed target state $x_T \in \mathcal{D}$ is an exponentially stable equilibrium of the closed-loop system.

The important observation is that the terminal cost~$\mathcal{T}$,
originally introduced as a soft penalty for the optimization
(Eq.~7), already possesses the mathematical structure of a CLF
for the droplet system. In error coordinates
$\tilde{\bm{x}} := \bm{x} - \bm{x}_T$, the terminal cost is a
quadratic form $\mathcal{T}(\tilde{\bm{x}}) = \tilde{\bm{x}}^{\!\top} P\,\tilde{\bm{x}}$
with $P = \mathrm{diag}(\mathcal{A}/X_T^2,\;\mathcal{B}/R_T^2)$,
which is positive definite and radially unbounded on a physical domain,
$\mathcal{D}:= \{(X, R) \in \R^2 : R > 0\}$. The CLF compatibility condition requires that,
for every state away from the target, there exists a control
input under which the combined rate of change of
$\mathcal{T}$ and the stage (viscous) cost $\mathcal{L}$ is non-positive \textbf{(SI Appendix)}
\setcounter{equation}{14}
\begin{align}
  \Phi^*(\bm{x})
  &:= \min_{G_0,\,\Delta G}
     \bigl\{\,
       \dot{\mathcal{T}}(\bm{x},\bm{G_0}, \bm{\Delta G})
       + \mathcal{L}(\bm{x},\bm{G_0}, \bm{\Delta G})
     \bigr\}
  \;\leq\; 0,\notag\\
  &\qquad \forall\;\bm{x}\in\mathcal{D}\setminus\{\bm{x}_T\}.
\end{align}
Since $\mathcal{L}$ is strictly convex in the controls (its
Hessian $\nabla_{\!\bm{U}}^2\mathcal{L} \succ 0$ on $\mathcal{D}$;
\textbf{see SI Appendix}), this minimization admits a unique closed-form
solution at each state. Evaluating it explicitly \textbf{(SI Appendix)}
shows that $\Phi^*(\bm{x}) \leq 0$ everywhere on $\mathcal{D}$,
except at $\bm{x} = \bm{x}_T$. This establishes
$\mathcal{T}$ as a valid, cost-compatible CLF
for the reduced-order droplet dynamics on~$\mathcal{D}$.

Since the reduced
dynamics~(9) are purely control-driven with no drift, and the viscous dissipation
cost is a positive-definite quadratic in the controls, the CLF
condition is not restricted to a linearized neighborhood of the
target. The droplet system's driftless, control-affine structure
ensures that $\mathcal{T}$ can always be decreased by a suitably
chosen control, regardless of how far the droplet is from
its target \textbf{(SI Appendix)}.
On any compact subset $\mathcal{K} \subset \mathcal{D}$ with
bounded droplet width $R \in [R_{\min}, R_{\max}]$, the
minimized decrease rate $\Phi^*$ satisfies a quadratic bound
in the error $\|\tilde{\bm{x}}\|^2$, meaning that the farther
the droplet is from its target in position or width, the faster
the CLF-compatible feedback corrects the deviation. This
quadratic decrease translates, via the comparison lemma, into
exponential convergence of $\mathcal{T}$ to zero: the terminal
cost decays as $e^{-ct/\bar{p}}$, where the rate $c/\bar{p}$ ( $\overline{p} = \lambda_{\max}(P)$ and $\lambda_{\max}$ is the largest eigen value of $P$)
depends explicitly on the viscosity $\mu$, the target state
$(X_T, R_T)$, and the terminal-cost weights $\mathcal{A}$,
$\mathcal{B}$ (derivation and numerical phase-portrait
verification in \textbf{SI Appendix}). Exponential stability ensures that the droplet reaches its
target state at a quantifiable rate rather than merely
approaching it asymptotically. Perturbations in initial
position or width are attenuated by a fixed fraction per
unit time, independent of their source. This is particularly relevant for
microfluidic systems where actuation noise, fabrication
tolerances, and thermal fluctuations continuously perturb the
droplet state during transport.

These results carry two practical implications for droplet
transport. 
First, the reduced-order system's inherent stabilizability, established through the CLF analysis, provides confidence that the TR and CTR strategies identified in Sections III and IV can be robustly implemented, since the underlying dynamics admit exponential convergence to any target state.
The terminal penalty introduced for optimization
purposes simultaneously provides stability guarantees, a
structural consequence of the driftless dynamics and the
physics-based dissipation cost. Second, since the decay rate
is given explicitly in terms of the microchannel parameters,
the terminal-cost weights can be selected a priori to
guarantee a desired convergence rate for a given transport
task, providing a quantitative design tool for microfluidic
control applications.

\section{Discussion}

The driftless, control-affine dynamics combined with a dissipation cost form a special problem class where PMP yields analytical controls and interpretable regime transitions. The TR and CTR strategies trade off
motion, deformation, and dissipation, providing a principled
framework for controlling externally actuated soft interfaces.
The PDE simulations confirm that these regimes persist in the
full continuum model across a wide range of capillary numbers,
with an optimal transport efficiency achieved at intermediate
$Ca_G$ where capillary restoring forces and viscous driving are balanced.
Standard linear controllers (proportional-integral-derivative, PID or Linear quadratic regulator, LQR) could regulate the droplet states around a setpoint but would not provide the physically interpretable TR and CTR transport strategies, which arise from the nonlinear coupling between droplet size and dissipation cost. An LQR design would linearize around the target and miss the energetic advantage of intermediate compaction that characterizes the CTR regime. Moreover, PID controllers do not minimize a cost, nor do they provide the formal stability guarantees established here through the CLF framework.

Our stability analysis reveals a structural consequence of this
problem class: the terminal cost $\mathcal{T}$, designed purely
as a performance penalty for finite-horizon optimization, is
simultaneously a valid CLF with a quadratic decrease rate \textbf{(SI Appendix)} and exponential stability.
We acknowledge that the use of terminal costs as CLFs has precedent in the
nonlinear MPC literature~\cite{jadbabaie2001thesis, mayne2000constrained}. The contribution
here is not the general CLF construction, but the demonstration that a
physically motivated dissipation cost derived from lubrication
hydrodynamics naturally produces a new problem structure for
which the CLF condition holds on~$\mathcal{D}$, rather than only
near the target. This property is atypical and arises directly from
the absence of autonomous drift in passive droplet transport.
For fluidic systems with no drift
term, a standard quadratic terminal penalty simultaneously
satisfies CLF conditions, eliminating the need for separate
Lyapunov construction. 
The TR and CTR regimes are therefore optimal, and the reduced-order system's inherent stabilizability ensures that robust feedback control to any target state is achievable.
This distinguishes the present
framework from existing microfluidic feedback strategies
which lack formal stability guarantees for the controlled droplet dynamics.

The analysis is established for the reduced-order
system~\eqref{eq:dynamics}, which captures the dominant slow
dynamics: position and shape modes evolve on the convective
scale  $\sim\mu R^4/|\ub|$ (where, $U \in [G_0,\,\Delta G]$), while higher-order spatial modes are damped by capillary relaxation on the faster scale $\sim\mu R^4/\gamma$. For capillary effects $O(1)$ or smaller, this separation justifies treating the ODE as the slow subsystem on an attracting manifold. 
While the ODE model captures intuitive droplet dynamics, shape stability requires explicit capillary effects ($\gamma$).
Establishing closed-loop stability
for the PDE requires constructing a control Lyapunov
functional on the Sobolev space $H^1(\Omega)$, which
measures not only the deviation of the height field $h(x,t)$
from the target profile but also the spatial gradient
$\partial_x h$, thereby penalizing interfacial roughness and
ensuring regularity of the controlled solution. This
infinite-dimensional extension, along with a rigorous
singular perturbation analysis linking the ODE and PDE
stability guarantees, is deferred to future work.
The present analysis assumes unconstrained controls. In physical microfluidic systems, pressure actuators have finite range. The CLF feedback shows that the required control remain moderate near the target.
For large deviations where saturation may arise, a constrained
model predictive control (MPC) formulation can be adopted while retaining terminal cost to preserve the stability guarantees.

\newpage
\section*{Code availability}
The complete source code for the ODEs, PDEs, and detailed finite element implementations will be released at

{\ttfamily\fontsize{9.73pt}{18pt}\selectfont\url{https://github.com/rajneeshanand/DropletOC}}
\section*{Acknowledgment}
We acknowledge Lehigh University’s High performance Computing resources. 

\bibliographystyle{IEEEtran} 
\bibliography{citation}

\end{document}


\pagestyle{empty}
\setlength{\headheight}{13.59999pt}

\date{\today}
\newpage
\section*{\LARGE Supplementary Information for}
\subsection*{\Large Optimal Control and Closed-Loop Stability of Droplet Transport in a Microchannel\\}

\author[\large Rajneesh Anand\textsuperscript{1} and Mayuresh V. Kothare\textsuperscript{1*}

*Corresponding Author: mvk2@lehigh.edu\\[4pt]
\normalsize\textsuperscript{1}Department of Chemical and Biomolecular Engineering, Lehigh University, Bethlehem, PA 18015, USA

\setcounter{equation}{0}
\renewcommand\theequation{S\arabic{equation}}

\newpage  
\section{Transport equations for the cost function}
\textbf{Depth-integrated continuity equation:} We start from the incompressible continuity equation $\frac{\partial \rho}{\partial t} + \nabla \cdot (\rho \mathbf{u}) = 0.$ For an incompressible fluid of constant density $\rho$, this reduces to $\nabla \cdot \mathbf{u} = 0.$ We consider a two-dimensional flow with velocity field $\mathbf{u} = (u(x,z,t),\, v(x,z,t)),$ where $u$ and $v$ represents the horizontal and vertical velocity, respectively. We considered droplet transport in a microchannel of fixed height $H$, bounded by rigid top and bottom walls. The droplet dynamics are modeled using a depth-averaged lubrication framework in which the scalar state variable $h(x,t)$ denotes the local thickness of the droplet phase measured from the bottom wall. Importantly, $h(x,t)$ describes the position of the droplet-carrier interface and may vary along the channel and does not represent deformation of the confining walls. The continuity equation becomes $\frac{\partial u}{\partial x} + \frac{\partial v}{\partial z} = 0.$ Integrate the continuity equation:
\begin{equation}
\int_0^{h(x,t)} 
\left(
\frac{\partial u}{\partial x}
+
\frac{\partial v}{\partial z}
\right) dz = 0.
\end{equation}
We applied the Leibniz rule for first term:
\[
\int_0^{h} \frac{\partial u}{\partial x} \, dz
=
\frac{\partial}{\partial x}
\left(
\int_0^{h} u \, dz
\right)
-
u(x,h,t)\frac{\partial h}{\partial x}.
\]

and, the second term:
\[
\int_0^{h} \frac{\partial v}{\partial z} \, dz
=
v(x,h,t) - v(x,0,t).
\]

Imposing impermeability at walls $v(x,0,t) = 0.$ Thus we have $\int_0^{h} \frac{\partial v}{\partial z} \, dz
=
v(x,h,t).$ We applied Kinematic boundary conditions, the free surface is described by the function $F(x,z,t) = z - h(x,t) = 0.$ Material conservation of the interface requires $\frac{DF}{Dt} = 0.$ Since $\frac{\partial F}{\partial t} = -\frac{\partial h}{\partial t},
\quad
\frac{\partial F}{\partial x} = -\frac{\partial h}{\partial x},
\quad
\frac{\partial F}{\partial z} = 1$.  We obtained  $v(x,h,t)
=
\frac{\partial h}{\partial t}
+
u(x,h,t)\frac{\partial h}{\partial x}.$ Substitute all terms back into the integrated continuity equation, we obtained $\frac{\partial}{\partial x}
\left(
\int_0^{h} u \, dz
\right)
-
u(x,h,t)\frac{\partial h}{\partial x}
+
\frac{\partial h}{\partial t}
+
u(x,h,t)\frac{\partial h}{\partial x}
= 0.$ The $u(x,h,t)\partial h/\partial x$ terms cancel, giving a final depth-integrated equation
\[
{\frac{\partial h}{\partial t}
+
\frac{\partial}{\partial x}
\left(
\int_0^{h} u \, dz
\right)
= 0.
}
\]
We have depth-averaged velocity $\langle u \rangle
=
\frac{1}{h}
\int_0^{h} u \, dz$ and the horizontal volumetric (mass) flux per unit width is $q = h \langle u \rangle = \int_0^{h} u \, dz.$ Thus, the depth-integrated continuity equation becomes
\begin{equation} \label{continuity}
{
\frac{\partial h}{\partial t}
+
\frac{\partial q}{\partial x}
= 0.
}
\end{equation}
Integrating over the domain $\frac{d}{dt}\int h(x,t)\,dx = 0$  expresses conservation of total fluid volume (mass).

\hfill $\blacksquare$

\textbf{Momentum balance in the lubrication limit:} At low Reynolds number, inertial forces are negligible and the flow is primarily governed by the Stokes equations $\nabla \cdot \boldsymbol{\sigma} = 0,$ where the stress tensor is $\boldsymbol{\sigma}
=
- p\,\mathbf{I}
+
\mu \left( \nabla \mathbf{u} + (\nabla \mathbf{u})^{T} \right).$ The velocity field is two-dimensional ${\textbf{u}} = (u(x,z,t),\, v(x,z,t)).$ The pressure is primarily capillary in origin $p = -\gamma \kappa,$ where the curvature $\kappa
=
\frac{\partial_{xx} h}
{\left[1 + (\partial_x h)^2 \right]^{3/2}}.$ Under the long-wave approximation, $\kappa \approx \partial_{xx} h$ has been assumed. We have an external pressure $p_{\text{ext}}$ as a control input for the flow, the effective pressure becomes $p(x,t) = -\gamma \partial_{xx} h + p_{\text{ext}}.$
The carrier oil is taken to be substantially more viscous than the droplet phase, so that the droplet-carrier interface at $z=h$ is effectively immobilized and a no-slip condition applies there.
For the lubrication scaling, let $H$ and $L$ denote the vertical and horizontal length scales, and define $\varepsilon = \frac{H}{L} \ll 1,$ with scalings $x = L\hat{x}, z = H\hat{z},
u = U\hat{u}, v = V\hat{v},$  incompressibility implies $V \sim \varepsilon U$ 
The $x$-momentum equation
\begin{equation}
-\frac{\partial p}{\partial x}
+ \mu \left(
\frac{\partial^2 u}{\partial z^2}
+
\frac{\partial^2 u}{\partial x^2}
\right) = 0.
\end{equation}
Since $\frac{\partial^2 u}{\partial x^2}
\sim
\varepsilon^2
\frac{\partial^2 u}{\partial z^2},$ the equation reduces to
$-\frac{\partial p}{\partial x}
+
\mu \frac{\partial^2 u}{\partial z^2}
= 0.$
Here,  $p_{\mathrm{ext}}(x,t)$ denote the externally imposed pressure (e.g.\ hydrostatic, magnetic, imposed axial pressure gradient, etc.).

The externally imposed forcing acts on the droplet through an effective head
that scales with the local droplet thickness, $p_{\mathrm{ext}}(x,t) = G(x,t)h(x,t)$,
as for a hydrostatic drive arising from the density contrast between the droplet
and the carrier phase. Here $G(x,t)$ is the axial driving field, with units of
force per unit volume (equivalently, of a pressure gradient).
From the lubrication approximation, the horizontal velocity profile is
$u(x,z,t) = \frac{z^2-hz}{2\mu}\frac{d}{dx}\!\left(\gamma h_{xx}+Gh\right).$
The depth-averaged volumetric flux per unit width is
$q(x,t) = \int_0^{h} u(x,z,t)\,dz.$ Evaluating the integral,

\begin{equation} \label{flux}
{
q(x,t)
=
-\frac{h^3}{12\mu}
\left(\gamma h_{xxx}+\partial_x(G(x,t)h)\right).
}
\end{equation} 

\hfill $\blacksquare$

\section{Reduced order model and ODE optimal control}

\textbf{ODE optimal transport objective:} The objective is to transport the droplet to a prescribed location with minimal energetic cost. The total cost functional is defined as $\mathcal{C} = \mathcal{W} + \mathcal{T},$ where $\mathcal{W}$ is the bulk viscous dissipation and $\mathcal{T}$ is a terminal penalty. The instantaneous dissipation rate is $\Phi = \mu \int (\nabla \mathbf{u}:\nabla \mathbf{u})\,dV.$ In the lubrication limit, the dominant contribution arises from vertical shear, $\Phi \approx \mu \int (u_z)^2\,dV.$ Using the lubrication velocity profile, we obtained
\begin{equation}
\int_0^h (u_z)^2 dz
=
\frac{h^3}{12\mu^2}
\left(\frac{d}{dx}\!\left(\gamma h_{xx}+Gh\right)\right)^2
\end{equation}

We can express this in terms of the flux $q$ yields $\mu$ $\int_0^h (u_z)^2 dz
=
\frac{12\mu}{h^3} q^2.$ The bulk cost is therefore
\begin{equation} \label{Wvisc}
{
\mathcal{W}
=
\int_0^T dt
\int 
 \left(\frac{12\mu}{h^3}\,q(x,t)^2\right) dx.
}
\end{equation}
The terminal cost penalizes deviation from a target position $x_T,$ and hence $T =\left[x_{\mathrm{com}}(T)-x_T\right]^2.$ The droplet center of mass is defined as
\begin{equation} \label{Xcom}
x_{\mathrm{com}}(t)
=
\frac{\int x\,h(x,t)\,dx}{\int h(x,t)\,dx}.
\end{equation}

The optimal transport problem is governed by a nonlinear PDEs for the film height
$h(x,t)$, which has infinitely many degrees of freedom. Moreover, standard
forward-in-time solution techniques are difficult because optimal control
problems require satisfying terminal conditions in addition to initial conditions. Two main approaches has been considered for the current study: Full PDEs approach which directly solve the PDE-constrained optimal control problem. While accurate, this approach is computationally expensive. Reduced-order model which project the PDE dynamics onto a finite set of low-dimensional modes, yielding a tractable system of ODEs.

\subsection{Reduced-Order Model}
For solving the full height field $h(x,t)$ droplet profile, we retain only a low order of modes which accounts for the shape, size, location of the droplet. Drop center-of-mass position, $X(t)$ and drop size or shape parameter, $R(t)$ have been adopted for state variables. This leads to a finite-dimensional dynamical system that can be analyzed using classical optimal control theory, in particular Pontryagin’s Maximum Principle. This approach is computationally efficient and provides clear physical insight into which modes are actively controlled. The control input is taken to be the axial driving field.

\begin{figure}
    \centering
    \includegraphics[width=0.5\linewidth]{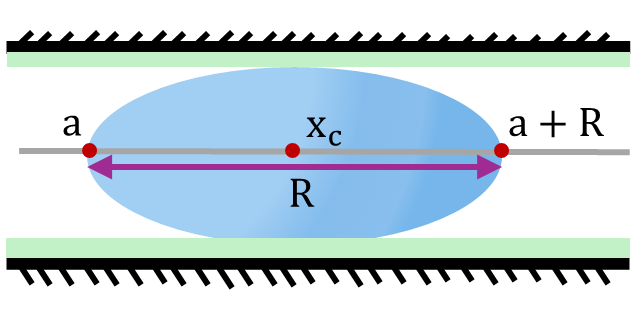}
    \caption{A schematic of the axisymmetric confined droplet geometry in a microchannel. The droplet is characterized by its axial center position $x_c$, left reference edge at $a$, and axial extent $R$, spanning from $a$ to $a+R$ along the channel centerline .}
    \label{fig:drop}
\end{figure}

For simplicity, we consider a minimal setting in which the spatial structure of the forcing is fixed. Accordingly, the control field is parameterized in a reduced form consisting of a mean component, $G_0(t)$ and a  spatial variation, $\Delta G(t)$. This parametrization yields a low-dimensional, physically interpretable control space suitable for optimal transport analysis. We assumed the droplet occupy a finite interval on the $x$-axis where left edge of the droplet $x = a(t)$, right edge of the droplet $x = a(t) + R(t)$, droplet midpoint (center of mass for symmetric shapes) $x_c(t) = a(t) + \frac{R(t)}{2}$, as shown in Fig.~\ref{fig:drop}.  To allow a simple spatial variation of the control input across the droplet, we assume a linear profile centered at the droplet midpoint.

\begin{equation} \label{G}
G(x,t)
=
G_0(t)
+
\Delta G(t)\,
\left(\frac{x - x_c(t)}{R(t)}\right)
\end{equation}
Since the droplet motion is passive, its dynamics result from the competition between viscous forcing and surface tension. The relative importance of these effects is quantified by a capillary number. The classical capillary number is $\mathrm{Ca} \sim \frac{\mu U}{\gamma}.$ For controlled passive transport, an analogous capillary number is introduced as
\begin{equation}
\mathrm{Ca}_G
\sim
\frac{\langle |\Delta G| \rangle\, R(t)^2}{\gamma}
\end{equation}
where $\langle |\Delta G| \rangle_T = (1/T) \int_0^T |\Delta G| dt$ denotes a characteristic or time-averaged value. $\mathrm{Ca}_G \ll 1$ refers surface tension dominates and the droplet remains nearly spherical and $\mathrm{Ca}_G \gg 1$ refers imposed pressure gradients strongly deform the droplet.

\textbf{Symmetric droplet profile:} The droplet surface is governed by capillarity, with Laplace pressure $p = -\gamma \kappa.$ If the internal pressure is uniform, then the curvature ($\kappa = \frac{d^2 h}{dx^2}$) is constant. Therefore, the droplet profile is parabolic $h(x,t) = c_2 x^2 + c_1 x + c_0$. We introduced the coordinate shifting $y = x - x_c(t)$, assuming symmetry about the center $\frac{dh}{dy}(0,t) = 0,$ and the height profile reduces to $h(y,t) = A(t) - B(t)\,y^2$, where $A(t) \text{and} B(t)$ is the maximum height at the droplet center and curvature coefficient, respectively. The boundary condition at the edges, $h\!\left(\pm \frac{R(t)}{2},t\right) = 0,$ implies $A(t) = \frac{B(t)R(t)^2}{4}.$ The droplet volume is conserved, $V = \int_{-R/2}^{R/2} h(y,t)\,dy.$ This calculation gives the droplet shape as we considered $V=1$

\begin{equation}\label{hxt}
h(x,t)
=
\frac{6}{R(t)^3}
\left(
\frac{R(t)^2}{4}
-
(x - x_c(t))^2
\right)
\end{equation}

Here we provide the Galerkin projection techniques to reduce PDEs to ODEs by assuming a parametrized shape for the solution (here, a parabolic droplet profile \ref{hxt}) depending on $X(t) \text{ and } R(t)$, substituting this form into the governing equations and projecting the residual dynamics onto a low order modes (moments of the height field). In this way, the first and second moments give the evolution of droplet position $X(t)$ and droplet size $R(t)$

The continuum dynamics of the droplet is governed by a height evolution equation \ref{hxt} that expresses mass conservation, with transport driven by a combination of surface tension and a spatiotemporally varying control \ref{G}. In the absence of control input, surface tension alone acts to relax the interface, and the resulting flux vanishes when the interface curvature is spatially uniform. Under this condition, the droplet naturally assumes a parabolic height profile. This observation motivates a reduced description in which the full height field is parametrized by only two time-dependent state variables: the droplet position, defined as the center of mass of the height distribution, and the droplet size, which is related to the second central moment (variance) of the profile.  The droplet position is defined as the center of mass of the height profile \ref{Xcom}. 

The variables $X(t)$ and $R(t)$ together capture all relevant features of the
droplet dynamics projected onto a two-dimensional reduced space where $X(t)$ describes droplet translation and $R(t)$ describes droplet deformation.
This constitutes a Galerkin projection of the full PDE dynamics onto the first
two moments of the height field.
The droplet motion is driven by the depth-averaged flux $q(x,t)$.
The rate of change of droplet position is obtained by integrating the flux \eqref{flux}:
\begin{equation}
\dot{X}(t)
=
\int_{X(t)-R(t)/2}^{X(t)+R(t)/2}
q(x,t)\,dx.
\end{equation}
To quantify the droplet width (or deformation), we introduce the second central
moment of the height profile $\Delta(t)
=
\int \big(x - X(t)\big)^2 h(x,t)\,dx.$ Substituting the parabolic profile and introducing \(u = x - X(t)\) and height field becomes $h(y,t)=\frac{6}{4R}
\left( 1- 
\frac{4u^2}{R^2} 
\right).$  We obtain 
\begin{equation}
\Delta  = \int_{-R/2}^{R/2}
u^2\,h(u)du 
\end{equation}
Hence, the droplet size is directly related to the second moment and we get $R = \sqrt{20\,\Delta}$.
We have $h = \frac{6}{R^3}\left(\frac{R^2}{4} - (x - X(t))^2\right) = \frac{3}{2R}\left(1 - \frac{4u^2}{R^2}\right)$ and
\begin{equation}
    q(u) = \frac{h^3(u)}{12\mu}\,\frac{d}{du}\bigl(G(u)\,h(u)\bigr) \nonumber
\end{equation}
Since the droplet shape is parabolic, $\frac{d^2h}{dx^2} = \text{const}$, which implies $\frac{d^3 h}{dx^3} = 0$, so the capillary contribution to the flux vanishes identically. This does not mean surface tension is physically absent, it acts on higher-order spatial modes that are projected out, providing a fast restoring force that keeps the droplet close to the parabolic manifold.
The imposed pressure gradient is linearized across the droplet $G(u) = G_0(t) + \Delta G(t)\,\frac{u}{R}$, 
where $G_0$ is the symmetric (uniform) part and $\Delta G$ is the antisymmetric part (left--right difference). Define $G_1 = \frac{\Delta G(t)}{R}$.
Now we compute $G(u)\,h(u) = \bigl(G_0 + G_1 u\bigr)\,\frac{3}{2R}\left(1 - \frac{4u^2}{R^2}\right)$ and its constituents
\begin{equation}
    \frac{d}{du}\bigl(G\,h\bigr) = \frac{3}{2R}\left[G_1\left(1 - \frac{4u^2}{R^2}\right) - \bigl(G_0 + G_1 u\bigr)\cdot\frac{8u}{R^2}\right] \nonumber
\end{equation}
\begin{equation}
    q = \frac{27}{64\mu R^4}\left(1 - \frac{4u^2}{R^2}\right)^3\left[G_1\left(1 - \frac{4u^2}{R^2}\right) - \bigl(G_0 + G_1 u\bigr)\frac{8u}{R^2}\right] \nonumber
\end{equation}
Since $G_0 \cdot 8u/R^2$ is an odd function integrated over the symmetric interval $[-R/2,\, R/2]$, its contribution to $\dot{X}$ vanishes. Thus:
\begin{equation}
    \dot{X} = \frac{27\,G_1}{64\,R^4\,\mu}\int_{-R/2}^{R/2}\left(1 - \frac{4u^2}{R^2}\right)^3\left[\left(1 - \frac{4u^2}{R^2}\right) - \frac{8u^2}{R^2}\right]du 
    =\frac{27\,G_1}{64\,R^4\,\mu}\int_{-R/2}^{R/2}\left[1 - \frac{24u^2}{R^2} + \frac{192u^4}{R^4} - \frac{640u^6}{R^6} + \frac{768u^8}{R^8}\right]du 
    \nonumber
\end{equation}
Recalling that $G_1 = \frac{\Delta G}{R}$, we obtain
\begin{equation}
    \dot{X} = \frac{9\,\Delta G}{70\,\mu\,R^4}
\end{equation}
The droplet translates only in response to the antisymmetric forcing $\Delta G$; a uniform field $G_0$ does not move the center of mass. The velocity scales as $R^{-4}$, so smaller droplets translate faster.
Since $R = \sqrt{20\Delta}$, we have
\begin{equation}
    \dot{R} = \frac{1}{2\sqrt{20\Delta}}\cdot 20\,\dot{\Delta} = \frac{10\,\dot{\Delta}}{R} 
    \text{  where  }
     \dot{\Delta} = 2\int_{x - R/2}^{x + R/2}(x - X(t))\,q(x,t)\,dx = 2\int_{-R/2}^{R/2} u\,q(u)\,du
    \nonumber
\end{equation}
Substituting the flux with $q = \frac{h^3}{12\mu}\cdot\frac{d}{du}(G\,h)$, we obtain
\begin{equation}
    q = \frac{h^3\,G_0}{12\mu}\,\frac{dh}{du}, \qquad h = \frac{3}{2R}\left(1 - \frac{4u^2}{R^2}\right), \qquad \frac{dh}{du} = \frac{3}{2R}\cdot\frac{-8u}{R^2} = \frac{-12u}{R^3} \nonumber
\end{equation}
since the odd-function argument now survives (because $u\cdot q$ picks out the $G_0$ term), we get
\begin{equation}
    \dot{\Delta} = -\frac{27\,G_0}{4\,\mu R^6}\int_{-R/2}^{R/2}\left(1 - \frac{4u^2}{R^2}\right)^3\,u^2\,du 
    = -\frac{27\,G_0}{4\,R^6}\cdot\frac{16R^3}{1260} = -\frac{3\,G_0}{35\,\mu\,R^3}
    \text{ and }
    \dot{R} = \frac{10\,\dot{\Delta}}{R} = -\frac{6\,G_0}{7\,\mu\,R^4}
    \nonumber
\end{equation}
The reduced-order ODE system governing droplet translation and deformation is:
\begin{align}
    \dot{X} = \frac{9\,\Delta G}{70\,\mu\,R^4}, \quad
    \dot{R} = -\frac{6\,G_0}{7\,\mu\,R^4}.
\end{align}
$\dot{X}$ is driven solely by the antisymmetric forcing $\Delta G$: a left--right pressure difference pushes the droplet. $\dot{R}$ is driven solely by the symmetric forcing $G_0$: a uniform body force compresses ($G_0 > 0$) or stretches ($G_0 < 0$) the droplet. The two controls $G_0$ and $\Delta G$ decouple into independent channels for deformation and translation, respectively. 
The Galerkin projection retains only the first two moments (position and variance) of the height field. Higher-order modes corresponding to front-back asymmetry and thinning near the contact line are governed by capillary relaxation and decay on the fast timescale. In the PDE simulations, the droplet remains intact for all parameter combinations studied ($Ca_G$ up to 1200, $X_T/R_T$ up to 0.3). No topological breakup events were observed; the failure mode at large $X_T/R_T$ is instead loss of film positivity near the contact line.

\hfill $\blacksquare$

\textbf{Power Dissipated:}
From the Stokes-flow velocity profile across the channel gap, the power dissipated per unit length is
\begin{equation}
    \int_0^h \mu\left(\frac{du}{dz}\right)^2 dz = \frac{12\mu}{h^3}\,q^2.
\end{equation}
Integrating over the droplet footprint:
\begin{equation}
    \mathcal{L} = \int_{x-R/2}^{x+R/2}\frac{12\mu}{h^3}\,q(x,t)^2\,dx = \frac{1}{12\mu}\int_{-R/2}^{R/2}h^3(u)\left(\frac{d}{du}(G\,h)\right)^2 du \nonumber
\end{equation}
With $G = G_0 + \frac{\Delta G}{R}\,u$ and $h = \frac{6}{R^3}\left(\frac{R^2}{4} - u^2\right)$, we compute:
\begin{equation}
    \mathcal{I} = \int_{-R/2}^{R/2}h^3\left(\frac{d}{du}(G\,h)\right)^2 du = \frac{7776}{R^{15}}\int_{-R/2}^{R/2}\left(\frac{R^2}{4} - u^2\right)^3\left(\frac{\Delta G\,R}{4} - 2G_0\,u - \frac{3\,\Delta G\,u^2}{R}\right)^2 du \nonumber
\end{equation}
Substitute $u = \frac{R}{2}\,y$, $y\in[-1,1]$, $du = \frac{R}{2}\,dy$:
    $\left(\frac{R^2}{4} - u^2\right)^3 = \frac{R^6}{64}(1-y^2)^3,$ 
The forcing term becomes
\begin{equation}
    \frac{\Delta G\,R}{4} - 2G_0\cdot\frac{R}{2}\,y - \frac{3\,\Delta G}{R}\cdot\frac{R^2}{4}\,y^2 = R\left[\frac{\Delta G}{4} - G_0\,y - \frac{3\,\Delta G}{4}\,y^2\right] \nonumber
\end{equation}
\begin{equation}
    \mathcal{I} = \frac{60.75}{R^6}\int_{-1}^{1}(1-y^2)^3\left(\frac{\Delta G}{4} - G_0\,y - \frac{3\,\Delta G}{4}\,y^2\right)^2 dy \nonumber
\end{equation}
\begin{equation}
    \mathcal{L}(t) = \frac{1}{12\mu}\cdot\mathcal{I} = \frac{81}{16\,\mu\,R^6}\int_{-1}^{1}(1-y^2)^3\left(\frac{\Delta G}{4} - G_0\,y - \frac{3\,\Delta G}{4}\,y^2\right)^2 dy \nonumber
\end{equation}

We define, $A(y) := \frac{\Delta G}{4} - G_0\,y - \frac{3\,\Delta G}{4}\,y^2.$ 
Since $(1-y^2)^3$ is even, all odd-power terms in $A(y)^2$ vanish upon integration. 
\begin{equation}
    A(y)^2 \bigg|_{\text{even}} = \frac{\Delta G^2}{16} + \left(G_0^2 - \frac{3\,\Delta G^2}{8}\right)y^2 + \frac{9\,\Delta G^2}{16}\,y^4 \nonumber
\end{equation}
We define the coefficients:
    $C_0 = \frac{\Delta G^2}{16}, C_2 = G_0^2 - \frac{3\,\Delta G^2}{8}, C_4 = \frac{9\,\Delta G^2}{16}.$
and use beta-function integrals $\int_{-1}^{1}y^{2m}(1-y^2)^3\,dy = 2\int_0^1 y^{2m}(1-y^2)^3\,dy$. 
Using the substitution $t = y^2$, $dt = 2y\,dy$, $dy = \frac{dt}{2\sqrt{t}}$:
\begin{equation}
    2\int_0^1 t^m(1-t)^3\,\frac{dt}{2\sqrt{t}} = \int_0^1 t^{(m-1)/2}(1-t)^3\,dt = B\!\left(\frac{m+\tfrac{1}{2}}{1},\;4\right) = \frac{\Gamma\!\left(\frac{m+1}{2}\right)\,\Gamma(4)}{\Gamma\!\left(m + \frac{9}{2}\right)} \nonumber
\end{equation}
Evaluating for each coefficient:

\noindent For $C_0$ ($m=0$):
\begin{equation}
    B\!\left(\tfrac{1}{2},\,4\right) = \frac{\Gamma(1/2)\,\Gamma(4)}{\Gamma(4.5)} = \frac{\sqrt{\pi}\cdot 6}{\tfrac{7}{2}\cdot\tfrac{5}{2}\cdot\tfrac{3}{2}\cdot\tfrac{1}{2}\sqrt{\pi}} \cdot \frac{6\sqrt{\pi}}{16\sqrt{\pi}/105} = \frac{32}{35} \nonumber
\end{equation}

\noindent For $C_2$ ($m=1$):
\begin{equation}
    B\!\left(\tfrac{3}{2},\,4\right) = \frac{\Gamma(3/2)\,\Gamma(4)}{\Gamma(5.5)} = \frac{\tfrac{1}{2}\sqrt{\pi}\cdot 6}{\tfrac{9}{2}\cdot\tfrac{7}{2}\cdot\tfrac{5}{2}\cdot\tfrac{3}{2}\cdot\tfrac{1}{2}\sqrt{\pi}} = \frac{32}{315} \nonumber
\end{equation}

\noindent For $C_4$ ($m=2$):
\begin{equation}
    B\!\left(\tfrac{5}{2},\,4\right) = \frac{\Gamma(5/2)\,\Gamma(4)}{\Gamma(6.5)} = \frac{\tfrac{3}{2}\cdot\tfrac{1}{2}\sqrt{\pi}\cdot 6}{\tfrac{11}{2}\cdot\tfrac{9}{2}\cdot\tfrac{7}{2}\cdot\tfrac{5}{2}\cdot\tfrac{3}{2}\cdot\tfrac{1}{2}\sqrt{\pi}} = \frac{96}{3465} \nonumber
\end{equation}

\begin{equation}
    \mathcal{L}(t) = \frac{81}{16\,\mu\,R^6}\left[C_0\cdot\frac{32}{35} + C_2\cdot\frac{32}{315} + C_4\cdot\frac{96}{3465}\right] \nonumber
\end{equation}

Coefficient of $G_0^2$: $\frac{81}{16\,\mu\,R^6}\cdot\frac{32}{315} = \frac{18}{35\,\mu\,R^6}$ and Coefficient of $\Delta G^2$ by collecting the $\Delta G^2$ pieces from $C_0$, $C_2$, and $C_4$: $ \frac{27}{154\,\mu\,R^6}.$
Therefore:
\begin{equation}\label{eq:lagrangian}
    \mathcal{L}(t) = \frac{1}{\mu\,R^6}\left[\frac{18}{35}\,G_0^2 + \frac{27}{154}\,\Delta G^2\right].
\end{equation}

The total cost (viscous dissipation) over the time horizon $[0,T]$ is $\mathcal{W} = \int_0^T \mathcal{L}\,dt$.
The running cost $\mathcal{L}$ is a positive-definite quadratic form in the controls $(G_0,\,\Delta G)$, weighted by $(\mu R^6)^{-1}$. This means that the cost of actuating the droplet grows steeply as the droplet shrinks ($R$ decreases). The state of the system is $(X,R)$; the controls are $(G_0,\,\Delta G)$; and $\mathcal{L} = \mathcal{L}(X,R,G_0,\Delta G)$.

\hfill $\blacksquare$

\subsection{Optimal control via Pontryagin's Maximum Principle}
The necessary components of the optimal control problem are:
\begin{enumerate}
    \item \textbf{State variable} $\psi(t) = \begin{bmatrix} X(t) \\ R(t) \end{bmatrix}$: things that evolve with time.
    \item \textbf{Control variable} $u(t) = (G_0(t),\,\Delta G(t))$: adjustable forcing fields.
    \item \textbf{Cost function} $\mathcal{J} = \int_0^T \mathcal{L}\,dt$: to be minimized.
    \item \textbf{Costates} $p(t) = \begin{bmatrix} p_X(t) \\ p_R(t) \end{bmatrix}$: Lagrange multipliers that enforce the dynamics.
\end{enumerate}

Here $p_X(t)$ is the costate conjugate to $X$ and $p_R(t)$ is the costate conjugate to $R$.
The Pontryagin Hamiltonian is $H(\psi,\,p,\,u) = p\cdot f(\psi,u) - \mathcal{L}(\psi,u)$, 
where $f$ collects the right-hand sides of the ODE system (\cite{controltheory}, \cite{Pontryagin}).
\begin{equation}
    H = p_X\left(\frac{9\,\Delta G}{70\,\mu\,R^4}\right) + p_R\left(\frac{-6\,G_0}{7\,\mu\,R^4}\right) - \frac{1}{\mu\,R^6}\left[\frac{18}{35}\,G_0^2 + \frac{27}{154}\,\Delta G^2\right] \nonumber
\end{equation}

\textbf{Necessary Conditions} Pontryagin's Maximum Principle (PMP) states: for optimal controls $(G_0^*,\,\Delta G^*)$, at each time $t$ the Hamiltonian must be maximized with respect to those controls:
\begin{equation}
    (G_0^*,\,\Delta G^*) = \arg\max_{G_0,\,\Delta G}\; H(X,\,R,\,p_X,\,p_R,\,G_0,\,\Delta G).
\end{equation}
Here, $G_0^*,\,\Delta G^*$ demonstrates the particular time-dependent forcing fields that minimize total dissipation.
Setting $\frac{\partial H}{\partial G_0} = 0$ and $\frac{\partial H}{\partial\Delta G} = 0$:
\begin{equation}
    -p_R\cdot\frac{6}{7\mu R^4} - \frac{2\cdot 18\,G_0}{\mu\,R^6\cdot 35} = 0 \qquad\Longrightarrow\qquad G_0^* = -\frac{5}{6}\,p_R\,R^2.
\end{equation}
\begin{equation}
    p_X\cdot\frac{9}{70\mu R^4} - \frac{27\cdot 2\,\Delta G}{154\,\mu\,R^6} = 0 \qquad\Longrightarrow\qquad \Delta G^* = \frac{11}{30}\,p_X\,R^2.
\end{equation}
The optimal symmetric forcing $G_0^*$ is proportional to $p_R$ (the ``shadow price'' of the droplet width), while the optimal antisymmetric forcing $\Delta G^*$ is proportional to $p_X$ (the shadow price of position). Each control is modulated by $R^2$, reflecting the droplet's size-dependent response.
Substituting the optimal controls back into $H$, we obtain conserved Hamiltonian, constant along optimal trajectories.
\begin{equation}
    H = \frac{1}{\mu\,R^2}\left[p_X\cdot\frac{9}{70}\cdot\frac{11}{30}\,p_X + p_R\cdot\frac{-6}{7}\cdot\left(-\frac{5}{6}\right)p_R\right] - \frac{1}{\mu\,R^6}\left[\frac{18}{35}\cdot\frac{25}{36}\,p_R^2 R^4 + \frac{27}{154}\cdot\frac{121}{900}\,p_X^2 R^4\right] \nonumber
\end{equation}
\begin{equation}
    H(\psi,\,p) = H(\psi,\,p,\,G_0^*,\,\Delta G^*) = \frac{1}{28\,\mu\,R^2}\left[10\,p_R^2 + \frac{33}{50}\,p_X^2\right] \equiv H_0 = \text{const}.
\end{equation}

Hamilton's equations for the states are $ \dot{X} = \frac{\partial H}{\partial p_X}, \dot{R} = \frac{\partial H}{\partial p_R}$.
Hamilton's equations for the costates are $\dot{p}_X = -\frac{\partial H}{\partial X}, \dot{p}_R = -\frac{\partial H}{\partial R}$.
Since $H$ has no explicit $X$-dependence, $\dot{p}_X = 0 \Longrightarrow p_X = p_0 = \text{constant}.$
This represents translational invariance. The dynamics depend only on the droplet's size and the controls, not on its absolute location. By Noether's theorem (or equivalently the Hamiltonian structure), this symmetry implies that the canonical momentum $p_X$ is a conserved quantity.

\textit{Remark:} Should the dissipation not fluctuate with position? Yes, the dissipation $\mathcal{L}$ does fluctuate with $R(t)$, $G_0(t)$, and $\Delta G(t)$. But $p_X$ is not the dissipation; it is the sensitivity of the cost to changes in the $X$-trajectory. Since $p_X$ is constant, we can treat it as a single parameter $p_0$ whose value is fixed by the boundary condition $X(T) = X_T$.

The ODE system with optimal controls is nonlinear due to the $R^{-4}$ factors. To linearize the costate dynamics and obtain an analytical solution, we reparametrize time.
Define a new time variable $\tau$ via $\dot{\tau} = \frac{d\tau}{dt} = \frac{1}{\mu\,R^2(t)}, \tau(0) = 0$. 
In this new time, the derivative transforms as $\frac{d}{dt} = \frac{1}{\mu R^2}\frac{d}{d\tau}$, so we define new velocities:
\begin{align}
    \mathcal{X}'(\tau) &= \dot{X}(t)\cdot\mu R^2 = \frac{\partial H}{\partial p_X}\cdot\mu R^2, \nonumber\\[4pt]
    \rho'(\tau) &= \dot{R}(t)\cdot\mu R^2 = \frac{\partial H}{\partial p_R}\cdot\mu R^2, \nonumber \\[4pt]
    p'(\tau) &= \dot{p}_R(t)\cdot\mu R^2 = -\frac{\partial H}{\partial R}\cdot\mu R^2 \nonumber
\end{align}
where primes denote $d/d\tau$, and we have introduced $\rho(\tau) \equiv R(t(\tau))$ and $p(\tau) \equiv p_R(t(\tau))$.
Since $p_X = p_0$ is constant, we also write $p_\circ \equiv p_R$. We have linearized equations in $\tau$-time:

Droplet position:
\begin{equation}
    \mathcal{X}'(\tau) = \mu R^2\cdot\frac{\partial H}{\partial p_X} = \mu R^2\cdot\frac{1}{28\mu R^2}\cdot\frac{2\cdot 33}{50}\,p_X = \frac{33\,p_0}{700}.
\end{equation}
Droplet width (radius):
\begin{equation}
    \rho'(\tau) = \mu R^2\cdot\frac{\partial H}{\partial p_R} = \mu R^2\cdot\frac{1}{28\mu R^2}\cdot 2\cdot 10\,p_R = \frac{5}{7}\,p.
\end{equation}
Costate $p_R$:
\begin{equation}
    p'(\tau) = \mu R^2\cdot\left(+\frac{2}{R}\right)\cdot\frac{1}{28\mu R^2}\left[10p_R^2 + \frac{33}{50}p_X^2\right] = 2\mu\,\rho\,H_0
\end{equation}

More directly, from $\rho' = \frac{5}{7}\,p$ we get $p = \frac{7}{5}\,\rho'$, and then $\rho''(\tau) = \frac{5}{7}\,p'(\tau) = \frac{5}{7}\cdot 2\mu\rho H_0 = \frac{10\,\mu\,H_0}{7}\,\rho$. 
We define $k^2 = \frac{10\,\mu\,H_0}{7}$
The width satisfies a linear second-order ODE: $\rho''(\tau) = k^2\,\rho(\tau)$.
The time reparametrization $\tau$ straightens out the nonlinear dynamics into a linear hyperbolic equation for $\rho$. The parameter $k$ encodes the conserved Hamiltonian (total energy of the optimal trajectory). Note that $s = k\tau(T)$ is dimensionless.
The general solution is $\rho(\tau) = A\cosh(k\tau) + B\sinh(k\tau)$.

Boundary conditions: $\rho(0) = R_0, \text{ and } \rho(\tau(T)) = R_T$

At $\tau = 0$: $\rho(0) = A\cdot 1 + B\cdot 0 = A$, so $A = R_0$.

At $\tau = \tau(T)$: $R_0\cosh(k\tau(T)) + B\sinh(k\tau(T)) = R_T$, so $B = \frac{R_T - R_0\cosh(k\tau(T))}{\sinh(k\tau(T))}$. 

Therefore,
\begin{equation}
      \rho(\tau) = R_0\cosh(k\tau) + \frac{R_T - R_0\cosh(k\tau(T))}{\sinh(k\tau(T))}\,\sinh(k\tau) \nonumber
\end{equation}
\begin{equation}
    \rho(\tau) = \frac{R_T\sinh(k\tau) + R_0\sinh\bigl(k(\tau(T) - \tau)\bigr)}{\sinh(k\tau(T))}.
\end{equation}
Solution for $\mathcal{X}(\tau)$ (Position):
Since $\mathcal{X}' = \frac{33\,p_0}{700}$ is constant:
\begin{equation}
    \mathcal{X}(\tau) = \frac{33\,p_0}{700}\,\tau \qquad\Longrightarrow\qquad X(T) = \mathcal{X}(\tau(T)) = \frac{33\,p_0}{700}\,\tau(T) \nonumber
\end{equation}
\begin{equation}
    \rho'(\tau) = \frac{1}{\sinh(k\tau(T))}\left[R_T\,k\cosh(k\tau) - R_0\,k\cosh\bigl(k(\tau(T) - \tau)\bigr)\right] \nonumber
\end{equation}
At $\tau = 0$:
\begin{equation}
    \rho'(0) = \frac{k\bigl(R_T - R_0\cosh(k\tau(T))\bigr)}{\sinh(k\tau(T))} \nonumber
\end{equation}
Since $\rho'(0) = \frac{5}{7}\,p(0)$:
\begin{equation}
    p(0) = \frac{7}{5}\,\rho'(0) = \frac{7\,k}{5}\cdot\frac{R_T - R_0\cosh(k\tau(T))}{\sinh(k\tau(T))}
\end{equation}

The conserved Hamiltonian evaluated at $\tau = 0$ gives
\begin{equation}
    H_0 = \frac{1}{28\,\mu\,R_0^2}\left[10\,p(0)^2 + \frac{33}{50}\,p_0^2\right].
\end{equation}
Using $k^2 = \frac{10\,\mu\,H_0}{7}$, we get $\mu\,H_0 = \frac{7\,k^2}{10}$, so:
\begin{equation}
    \frac{7\,k^2}{10} = \frac{1}{28\,R_0^2}\left[10\,p(0)^2 + \frac{33}{50}\,p_0^2\right].
\end{equation}
Substituting the expressions for $p(0)$ and $p_0$ and we define $s = k\tau(T)$:

(a) For $p(0)$:
\begin{equation}
    10\,p(0)^2 = 10\cdot\frac{49\,k^2}{25}\cdot\frac{(R_T - R_0\cosh s)^2}{\sinh^2 s} = \frac{98\,k^2\,(R_T - R_0\cosh s)^2}{5\,\sinh^2 s}, \nonumber
\end{equation}
(b) For $p_0$: From $\mathcal{X}' = 33p_0/700$ and $X_T = (33p_0/700)\tau(T)$:
\begin{equation}
    \frac{33}{50}\,p_0^2 = \frac{33}{50}\cdot\frac{700^2\,k^2\,X_T^2}{33^2\,s^2} = \frac{9800\,k^2\,X_T^2}{33\,s^2}. \nonumber
\end{equation}
Substituting into the Hamiltonian conservation equation and dividing both sides by $k^2$ and multiplying by $28\,R_0^2$: and after rearranging, we obtain
\begin{equation}
    \frac{9800}{33\,s^2}\,X_T^2 = \frac{98\,R_0^2}{5} - \frac{98\,(R_T - R_0\cosh s)^2}{5\,\sinh^2 s} 
\end{equation}
Normalizing by $R_0^2$ and defining $\sigma = R_T/R_0$:
\begin{equation}
    \frac{9800}{33\,s^2}\left(\frac{X_T}{R_0}\right)^2 = \frac{98}{5}\left(1 - \frac{(\sigma - \cosh s)^2}{\sinh^2 s}\right) \nonumber
\end{equation}
Therefore:
\begin{equation}
    \left(\frac{X_T}{R_0}\right)^2 = \frac{33\,s^2}{500}\cdot\frac{-\sigma^2 + 2\sigma\cosh s - 1}{\sinh^2 s},
\end{equation}
which gives the fundamental implicit relation:
\begin{equation}
    \frac{X_T}{R_0} = \sqrt{\frac{33\,s^2}{500\,\sinh^2(s)}\left[\frac{R_T}{R_0}\bigl(2\cosh(s) - \frac{R_T}{R_0}\bigr) - 1\right]},
\end{equation}

Given initial and target droplet states $(R_0, R_T, X_T)$, the equation implicitly determines the dimensionless parameter $s$, which in turn fixes the entire optimal trajectory and control policy through the analytical solutions derived above.
It is important to note that the expression under the square root is not always positive: for certain combinations of $X_T$ (too large a translation) and $R_T/R_0$ (too extreme a size change), no real $s > 0$ exists. In such cases, there is no continuous optimal policy, the droplet cannot reach the target in a smooth, dissipation-minimizing manner.
This does not contradict controllability. Controllability requires the existence of any path (not necessarily optimal or smooth) that gets the droplet from state A to state B. Here, system is controllable. But the optimal control problem imposes additional structure (minimizing dissipation), and for sufficiently demanding targets, the optimal cost diverges or the optimal trajectory ceases to exist in the classical sense.
\hfill $\blacksquare$

\section{PDE optimal control}

This sections presents the governing equations and numerical formulation for the full nonlinear PDE optimal control problem. The complete derivation of the weak form, the semi-implicit time-stepping strategy (Adams-Bashforth (AB2)/Crank-Nicolson (CN) splitting), the FEniCS (\cite{Fenics}) implementation using \texttt{pycma} - a python package (\cite{Hansen2019pycma}), and the CMA-ES optimization workflow for the current work are documented in the project repository at \url{https://github.com/rajneeshanand/DropletOC}. 
The total cost driving the CMA-ES (\cite{CMAES}) optimization comprises three terms $\mathcal{J} = \mathcal{W} + \mathcal{T} + \mathcal{R}$.
\begin{equation}\label{visc_cost}
\mathcal{W}
=
\int_{0}^{T} dt
\int dx\;
\frac{h^{3}}{12\mu}
\left[
\partial_{x}(Gh + \gamma \partial_{xx}h - \Pi
\right)]^{2}
\end{equation}
\begin{equation}\label{terminal_cost}
    \mathcal{T} = \mathcal{A}\left(\frac{X(T) - X_T}{X_T}\right)^2 + \mathcal{B}\left(\frac{R(T) - R_T}{R_T}\right)^2,
\end{equation}
with $\mathcal{A} = \mathcal{B}= 10^3$. The large penalty weights enforce the target state as a soft constraint: the optimizer is strongly incentivized to reach $(X_T, R_T)$ but retains the flexibility to trade a small endpoint error against a large dissipation saving.
\begin{equation}
\mathcal{R}
=
\frac{\alpha}{N_{c}}
\int_{0}^{T} dt
\left[
(\partial_{t}G_{0})^{2}
+
(\partial_{t}G_{1})^{2}
\right]
\end{equation}
where $N_c = 100$ is the number of control time points and $\alpha$ is a regularization weight. Since the control is piecewise linear, $\mathcal{R} \to 0$ as $N_c \to \infty$ for fixed $\alpha$ and bounded time derivatives. The regularization term penalizes rapid switching, selecting smoother control protocols that are physically realizable.
The three cost terms encode a hierarchy of objectives: $\mathcal{T}$ ensures the droplet arrives at the target, $\mathcal{W}$ penalizes wasteful forcing, and $\mathcal{R}$ (\cite{brocket}) prevents unphysical high-frequency control chatter. The relative weighting determines the Pareto trade-off between precision and efficiency.
A thin precursor film of thickness $\delta = 10^{-2}$ (\cite{BrochardWyart1991},\cite{deGennes2013}, \cite{Pahlavan2018}) coats the substrate ahead of the droplet, regularizing the contact-line singularity. The disjoining pressure (\cite{deGennes1985}) enforces the equilibrium contact angle $\phi_0 = \pi/4$:
\begin{equation}
    \Pi(h) = \frac{\mathcal{P}}{h^3}\left(1 - \frac{\delta}{h}\right), \qquad \Pi'(h) = \frac{\mathcal{P}}{h^5}(4\delta - 3h),
\end{equation}
where $\mathcal{P} = 3\gamma\delta^2\tan^2\phi_0$. This form is equivalent to $6\gamma\delta^2(1 - \cos\phi_0)$ for small angles but performs better numerically at finite $\phi_0$.
The disjoining pressure creates a repulsive barrier near $h = \delta$ that prevents the film from thinning to zero, while simultaneously selecting the macroscopic contact angle. It is evaluated explicitly via AB2 because it is strongly nonlinear and localized near the contact line, treating it implicitly would couple every degree of freedom to the contact-line region and destroy the banded structure of the linear system.
The initial droplet is a symmetric, sessile drop centered at the origin with equilibrium size $R_0 = \sqrt{6}/\tan\phi_0 = \sqrt{6}$:
\begin{equation}
    h(x, 0) = \begin{cases}
        \delta + \dfrac{6}{R_0^3}\left[1 - R_0\delta\right]\left(\dfrac{R_0^2}{4} - x^2\right), & x \in \left[-\tfrac{R_0}{2},\,\tfrac{R_0}{2}\right], \\[8pt]
        \delta, & \text{otherwise}.
    \end{cases}
\end{equation}
This profile satisfies volume conservation ($\int h\,dx = 1$) and is the stable steady state of the passive dynamics when the activity is turned off. The surface tension is varied over the range $\gamma = 0.1$--$1.0$.
The simulation hyperparameters are defined in the Github page. The convergence criterion is set as
\begin{equation}
    \Delta\mathcal{C} \leq \epsilon_{\text{tol}}\left(\mathcal{C}_{\text{median}}^0 - \mathcal{C}_{\text{median}}^{\min}\right),
\end{equation}
where $\mathcal{C}_{\text{median}}^0$ is the initial median fitness, $\mathcal{C}_{\text{median}}^{\min}$ is the best median encountered, and $\epsilon_{\text{tol}} = 10^{-4}$. The maximum number of CMA-ES iterations is capped at $M_{\text{iter}} = 8 \times 10^4$.

\section{Transport efficiency}
To calculate transport efficiency, we define minimum energy that must be dissipated during droplet transport, we consider a height field $h(x,t)\ge 0$ with unit mass, evolving by Eq.~\eqref{continuity} and Eq.~\eqref{flux}. The total viscous dissipation in the drop
\begin{equation}
\begin{aligned}\label{cost} 
\mathcal{W}
&=
\int_{0}^{T} dt
\int \frac{h^{3}}{12\mu}
(\partial_{x}\sigma)^{2} dx
=
12\mu
\int_{0}^{T} dt
\int \frac{q^{2}}{h^{3}} dx
\end{aligned}
\end{equation}

The global translation $X(t)=\int x h(x,t)\,dx$ satisfies $\dot{X}=\int q\,dx$, and the width,
$\Delta(t)=\int (x-X)^{2} h\,dx$ follows $\Delta = 2\int (x-X) q\,dx$.
Using Cauchy–Schwarz on the integral of $\dot{X}$

\begin{equation}
|X(T)|^{2}
=
\left|
\int_{0}^{T} dt
\int \frac{q}{h} h\,dx
\right|^{2}
\le
\left(
\int_{t,x}
\frac{q^{2}}{h^{2}}
\right)
\left(
\int_{t,x} h^{2}
\right)
\end{equation}
Hölder’s inequality (Eq.~\eqref{Holders}) sets the upper bounds of droplet height using
$\|h\|_{\infty}=\sup_{x} h(x,t)$, we use Eq.\eqref{cost}, to build Eq.~\eqref{A4}, and combining these results gives the displacement bound Eq.~\eqref{A5}
\begin{equation}\label{Holders}
\int_{0}^{T} dt
\int h^{2}\,dx
\le
\|h\|_{\infty}
\int_{0}^{T} dt
\int h\,dx
=
T\|h\|_{\infty}
\end{equation}
\begin{equation}\label{A4}
\begin{aligned} \int_{0}^{T} dt
\int \frac{q^{2}}{h^{2}} & dx
\le 
\|h\|_{\infty}
\int_{0}^{T} dt
\int \frac{q^{2}}{h^{3}} dx
=
\frac{\|h\|_{\infty}\mathcal{W}}{12\mu}
\end{aligned}
\end{equation}
\begin{equation}\label{A5}
X(T)^{2}
\le
\frac{\|h\|_{\infty}^{2} T\,\mathcal{W}}{12\mu}
\end{equation}
\noindent
We evaluate the variance using Cauchy–Schwarz inequality (\ref{A6}) and Hölder’s inequality (\ref{A7}). This gives (\ref{A8})
\begin{equation}\label{A6}
|\Delta(T) - \Delta(0)|^{2}
= 
4\left|\int_{0}^{T} dt \int q (x - X)\,dx\right|^{2}
\end{equation}
\begin{equation}\label{A7}
\begin{aligned}|\Delta(T) - \Delta(0)|^{2}
&=
\left|
\int_{0}^{T} dt \int h^{2} (x - X)^{2}\,dx
\right|
\le 
\|h\|_{\infty}
\int_{0}^{T} \Delta(t)\,dt
=
T \|h\|_{\infty}\,\langle \Delta \rangle_{T}
\end{aligned}
\end{equation}
\begin{equation}\label{A8}
|\Delta(T) - \Delta(0)|^{2}
\le
\frac{4\|h\|_{\infty}^{2} T \mathcal{W}}{12\mu}\,
\langle \Delta \rangle_{T}
\end{equation}

\noindent
$\langle \Delta \rangle_{T}$ is the time-average of $\Delta(t)$. Combining these results gives a minimum energetic cost (Eq.~\eqref{minimum_cost}).
As optimal control solution must satisfy $\mathcal{W}_{\mathrm{opt}} \ge \mathcal{W}_{\min}$, we define the efficiency (Eq.~\eqref{eff}) to quantify the transport energy cost.
\begin{equation}\label{minimum_cost}
\mathcal{W}_{\min}
=
\frac{6\mu}{T h_{\infty}^{2}}
\left[
X(T)^{2}
+
\frac{|\Delta(T) - \Delta(0)|^{2}}{4\langle \Delta \rangle_{T}}
\right]
\le
\mathcal{W}
\end{equation}
\begin{equation}\label{eff}
\eta
=
\frac{\mathcal{W}_{\min}}{\mathcal{W}_{\mathrm{opt}}}
\le 1
\end{equation}

\hfill $\blacksquare$

\section{Control Lyapunov Function-Based Stability Analysis for ODE}

\textbf{Note:} Throughout, $\norm{\cdot}$ denotes the Euclidean norm on $\R^n$. For a $C^1$ function $\Phi:\R^n\to\R$, we write $\nabla\Phi(\xb)$ for its gradient (as a row vector) and $D\Phi(\xb)$ for the same object when we wish to emphasize the Fr\'echet derivative. For a positive-definite matrix $P\in\R^{n\times n}$, we write $\lambda_{\min}(P)$ and $\lambda_{\max}(P)$ for its smallest and largest eigenvalues. Sublevel sets of a function $V$ are denoted $\Omega_r := \{\xb : V(\xb) \leq r^2\}$.

\begin{definition}[ODE model]
The reduced-order droplet dynamics are given by the control-affine system
\begin{equation}\label{eq:dynamics}
    \dot{\xb} = \fb(\xb, U) := \begin{pmatrix} \dfrac{9\,\Delta G}{70\,\mu\, R^4} \\[10pt] -\dfrac{6\,G_0}{7\,\mu\, R^4} \end{pmatrix}
\end{equation}
where $\mu \text{ (fluid viscosity)} > 0$. Observe that $f(x, 0)=0,\forall x$, it depicts that the droplet is stationary in the absence of external forcing.
The dynamics has no surface tension ($\gamma$) or drift terms, the droplet moves only when actively controlled. This property (absence of autonomous drift) promotes controllability and will be exploited in the stability analysis.
\end{definition}

\subsection{Controllability}\label{sec:controllability}
Before discussing stability, we must verify that the system is controllable, i.e., one can control the droplet from any initial state ($t=0$) to the final state ($t=T$).

\begin{proposition}[Controllability]\label{prop:controllability}
The reduced-order system~\eqref{eq:dynamics} is controllable on the domain $\mathcal{D} := \{(X, R) \in \R^2 : R > 0\}$.
\end{proposition}

\begin{proof}
We apply the Lie algebra rank condition (LARC), also known as the Chow--Rashevskii theorem (see Ref.~\cite{ref15} for details). The system~(\eqref{eq:dynamics}) is control-affine $\dot{\xb} = g_1(\xb)\,G_0 + g_2(\xb)\,\Delta G$, with
$g_1(\xb) = (0,\; -6/(7\mu R^4))^T$ and $g_2(\xb) = (9/(70\mu R^4),\; 0)^T$.
Then $\det[g_1\; g_2] = 27/(245\mu^2 R^8) \neq 0$ on $\mathcal{D}$. Since $\text{rank}\{g_1, g_2\} = 2 = \dim(\xb)$ everywhere, the Lie algebra rank condition (LARC)~\cite{ref15} is satisfied, implying controllability on $\mathcal{D}$.
\end{proof}

\textbf{Remark:} (Sufficiency of LARC for driftless systems)
For driftless control-affine systems of the form $\dot{\bm{x}} = \sum_{i} g_i(\bm{x})\, U_i$, the Lie algebra rank condition (LARC) is both necessary and sufficient for small-time local controllability at every point where the rank condition holds~\cite{controltheory}. Since our system (Eq.~\eqref{eq:dynamics}) has exactly two states and two independent input vector fields with $\mathrm{rank}\{g_1, g_2\} = 2$ everywhere on $\mathcal{D}$, no additional Lie brackets are needed. The first-order fields already span the tangent space. Controllability on $\mathcal{D}$ then follows because any two points in $\mathcal{D}$ can be connected by a piecewise-constant control trajectory that remains in $\mathcal{D}$, since the vector fields $g_1$ and $g_2$ point along the coordinate directions $(0,\,\cdot\,)^T$ and $(\,\cdot\,,0)^T$ respectively, and no trajectory generated by bounded controls in finite time can reach $R = 0$ from $R > 0$. To see this, note that $\dot{R} = -6\,G_0/(7\,\mu\,R^4)$, so driving $R \to 0$ requires $|G_0| \to \infty$ or $t \to \infty$, neither of which is permitted under bounded controls and finite horizons. The boundary $R = 0$ is therefore not reachable from $\mathcal{D}$, and the domain $\mathcal{D}$ is positively invariant under bounded controls.

\subsection{Stage cost}\label{sec:stage_cost}

The stage cost (also called the incremental or running cost) is the instantaneous viscous dissipation rate in the droplet, as shown in Eq. \eqref{eq:lagrangian}.

\begin{proposition}[Convexity]\label{prop:stage_pos_def}
For every fixed $\xb \in \mathcal{D}$: (a) $\mathcal{L}(\xb, U) \geq 0$; (b) $\mathcal{L}(\xb, U) = 0$ iff $U = \bm{0}$; (c) $\mathcal{L}$ is strictly convex in $U$.
\end{proposition}

\begin{proof}
Parts (a)--(b) follow from $(\mu R^6)^{-1} > 0$ on $\mathcal{D}$. For (c), the Hessian
$\nabla^2_{U}\mathcal{L} = \frac{2}{\mu R^6}\text{diag}(18/35,\; 27/154)$
is positive definite on $\mathcal{D}$, establishing strict convexity.
\end{proof}

\noindent
\textbf{Remark:}
The stage cost is not positive definite in $\xb$. It vanishes for all $x$ when $U = 0$. This reflects that a stationary droplet (with no applied control) dissipates no energy, regardless of the states. This distinguishes our problem from classical regulation problems where the stage cost typically penalizes both state deviation and control effort. The state-independent nature of the stage cost places particular importance on the terminal cost for establishing stability, as we shall see later.

\subsection{Terminal cost}\label{sec:terminal_cost}   

The terminal cost (Eq.\eqref{terminal_cost}) penalizes the deviation of the final state from the target,
where $\mathcal{A}, \mathcal{B} > 0$ are positive weighting coefficients for the terminal cost. In error coordinates $\tilde{\xb} := (\tilde{X}, \tilde{R})^T$ where $\tilde{X} := X(t) -X_T,$  and $\tilde{R} := R(t)-R_T$
\begin{equation}\label{eq:terminal_error}
    \mathcal{T}(\tilde{\xb}) = \tilde{\xb}^T P\, \tilde{\xb}, \qquad P = \begin{pmatrix} \mathcal{A}/X_T^2 & 0 \\ 0 & \mathcal{B}/R_T^2 \end{pmatrix}
\end{equation}

\begin{proposition}[Positive Definiteness]\label{prop:terminal_pos_def}
$\mathcal{T}(\tilde{\xb})$ satisfies: (a) $\mathcal{T} \geq 0$; (b) $\mathcal{T} = 0$ iff $\tilde{\xb} = \bm{0}$; (c) radial unboundedness; and (d) quadratic bounds
\begin{equation}\label{eq:quadratic_bounds}
\underline{p}\,\norm{\tilde{\xb}}^2 \leq \mathcal{T}(\tilde{\xb}) \leq \overline{p}\,\norm{\tilde{\xb}}^2
\end{equation}
where $\underline{p} = \lambda_{\min}(P)$ and $\overline{p} = \lambda_{\max}(P)$.
\end{proposition}

\begin{proof}
Parts (a)--(c) follow immediately from the fact that $P$ is a real, symmetric, PD matrix ($\mathcal{A}, \mathcal{B}, X_T, R_T > 0$). For part (d), the Rayleigh quotient characterization of eigenvalues gives
    $\lambda_{\min}(P)\,\norm{\tilde{\xb}}^2 \leq \tilde{\xb}^T P\, \tilde{\xb} \leq \lambda_{\max}(P)\,\norm{\tilde{\xb}}^2 \forall\, \tilde{\xb} \in \R^2$,
which is exactly Eq.\eqref{eq:quadratic_bounds}.
\end{proof}

\begin{definition}[Finite-horizon cost function]\label{def:total_cost}
Given a horizon length $T > 0$, initial state $\xb_0$, and admissible control trajectory $U(\cdot) \in L^2([0,T]; \R^2)$, the finite-horizon cost is
\begin{equation}\label{eq:total_cost}
    J_T(\xb_0, \ub(\cdot)) := \underbrace{\int_0^T \mathcal{L}(\xb(t), U(t))\, dt}_{\text{Running cost } \mathcal{W}} + \underbrace{\mathcal{T}(\xb(T))}_{\text{Terminal cost}},
\end{equation}
$x(t)$ satisfies~(Eq.\eqref{eq:dynamics}) with $x(0)=x_0$. The optimal value function is
\begin{equation}\label{eq:value_function}
    J_T^*(\xb_0) := \inf_{U(\cdot)} J_T(\xb_0, U(\cdot)).
\end{equation}
\end{definition}

\subsection{The CLF Framework}\label{sec:CLF_background}

In this section, we list the key results from control Lyapunov functions~\cite{stability1} that form the theoretical backbone of our stability analysis.

\begin{definition}[Control Lyapunov Function]\label{def:CLF}
Consider a general nonlinear system $\dot{x} = f(x, u)$ with equilibrium at the origin. A $C^1$, proper, PD function $V: \R^n \to \R_+$ is called a control Lyapunov function (CLF) if
\begin{equation}\label{eq:CLF_def}
    \inf_{U \in \R^m} \left[ DV(\xb) \cdot \fb(\xb, U) \right] < 0 \quad \forall\, \xb \neq \bm{0}.
\end{equation}
\end{definition}

\begin{definition}[Cost-compatible CLF]\label{def:compatible_CLF}
A control Lyapunov function $V$ is said to be compatible with the incremental cost $\mathcal{L}(x, u)$ if there exists a neighborhood $\mathcal{N}$ of the origin such that
\begin{equation}\label{eq:compatible_CLF}
    \min_{U} \left\{ DV(\xb) \cdot \fb(\xb, U) + \mathcal{L}(\xb, U) \right\} \leq 0 \quad \forall\, \xb \in \mathcal{N}.
\end{equation}
\end{definition}

\subsection{Standing properties}\label{sec:assumptions}
Here, we state required properties and verify it to prove the main stability theorems.

\noindent
\textbf{Property 1:} [Domain restriction]\label{ass:domain}
The state evolves in the physical domain $\mathcal{D} = \{(X, R) \in \R^2 : R > 0\}$. All subsequent analysis is restricted to a compact subset $\mathcal{K} \subset \mathcal{D}$ containing the target state $\xb_T = (X_T, R_T)$ in its interior, with $R_{\min} \leq R \leq R_{\max}$ for some $0 < R_{\min} < R_T < R_{\max} < \infty$. Controls are taken in $U(·) \in L^2([0,T]; \R^2)$; no pointwise bound is imposed.

\noindent
\textbf{Property 2:} [Regularity]\label{ass:regularity}
The vector field $f(x,u)$ defined in~(Eq.\eqref{eq:dynamics}) is $C^{\infty}$ on $\mathcal{D} \times \R^2$. The stage cost $\mathcal{L}(x,U)$ and terminal cost $\mathcal{T}(x)$ are $C^{\infty}$ on $\mathcal{D} \times \R^2$ and $\mathcal{D}$, respectively.

\noindent
These regularity conditions are trivially satisfied: the dynamics~(Eq.\eqref{eq:dynamics}) are rational functions of $(X, R, G_0, \Delta G)$ with the only singularity at $R = 0$, which is excluded by Property-1. The costs are polynomial (hence $C^{\infty}$) on $\mathcal{D}$.


\noindent
\textbf{Property 3:} [Existence of optimal trajectories]\label{ass:existence}
For each $\xb_0 \in \mathcal{K}$ and $T > 0$, the minimum in $J_T^*(\xb_0) = \inf_{U(\cdot)} J_T(\xb_0, U(\cdot))$ is attained by some admissible control $U_T^*(\cdot; \xb_0)$.

\noindent
The conditions require (i) coercivity of the running cost in the control, 
(ii) the droplet dynamics are continuous and the set $f(x, \R^m)$ is convex 
for all $x$ (since $\fb$ is affine in $U$), and (iii) lower semicontinuity of 
the cost. For (i), $\mathcal{L}$ grows quadratically in $\norm{U}$ on 
$\mathcal{K}$, so minimizing sequences are bounded in $L^2([0,T]; \R^2)$; this 
coercivity replaces compactness of the control set, which is not assumed here. 
For (iii), $\mathcal{L}$ is continuous on $\mathcal{D} \times \R^2$ and convex 
in $U$ (Proposition~\ref{prop:stage_pos_def}), hence weakly lower semicontinuous 
along the minimizing sequence. The terminal cost is continuous, so the infimum 
is attained.

\subsection{Terminal cost $\mathcal{T}$ as a control Lyapunov function}
We now establish the terminal cost $\mathcal{T}$ satisfies the CLF compatibility condition for the droplet transport system.

\begin{theorem}[$\mathcal{T}$ as a cost-compatible CLF]\label{thm:CLF_main}
Let $\mathcal{T}(\xb)$ be the terminal cost defined in~(Eq.\eqref{terminal_cost}), and let $\mathcal{L}(\xb, U)$ be the stage cost defined in~(Eq.\eqref{eq:lagrangian}). Then:
\begin{enumerate}[label=(\alph*)]
    \item $\mathcal{T}$ is a CLF for the system on $\mathcal{D} \setminus \{\xb_T\}$.
    \item $\mathcal{T}$ is compatible with $\mathcal{L}$ in the sense that there exists a neighborhood $\mathcal{N}$ of $\xb_T$ such that
    \begin{align}\label{eq:CLF_compat_droplet}
        \min_{U \in \R^2} \left\{ \nabla \mathcal{T}(\xb) \cdot \fb(\xb, U) + \mathcal{L}(\xb, U) \right\} \leq 0 \quad  \forall\, \xb \in \mathcal{N} \setminus \{\xb_T\}\nonumber
    \end{align}
\end{enumerate}
\end{theorem}

\begin{proof}
We proceed in several steps.

\noindent Step 1: Compute the gradient of $\mathcal{T}$
From~\eqref{terminal_cost}:
\[\label{eq:grad_T}
    \nabla \mathcal{T} = \left(\pd{\mathcal{T}}{X},\; \pd{\mathcal{T}}{R}\right) = \left(\frac{2\mathcal{A}(X - X_T)}{X_T^2},\; \frac{2\mathcal{B}(R - R_T)}{R_T^2}\right)
\]

\noindent Step 2: Compute the Lie derivative $\dot{\mathcal{T}} = \nabla\mathcal{T} \cdot \fb$.
\begin{align}
    \dot{\mathcal{T}}(\xb, U) = \nabla \mathcal{T} \cdot \fb(\xb, U) 
    = \frac{9\mathcal{A}\,\tilde{X}}{35\,\mu\, X_T^2\, R^4}\,\Delta G - \frac{12\mathcal{B}\,\tilde{R}}{7\,\mu\, R_T^2\, R^4}\, G_0 \label{eq:Tdot}
\end{align}

\noindent Step 3: Minimize $\Phi(\xb, U) := \dot{\mathcal{T}}(\xb, U) + \mathcal{L}(\xb, U)$ over $U$.
\begin{align}\label{eq:Phi}
    \Phi(\xb, U) = \underbrace{\frac{9\mathcal{A}\,\tilde{X}}{35\,\mu\, X_T^2\, R^4}}_{=:\, \alpha(\xb)}\,\Delta G - \underbrace{\frac{12\mathcal{B}\,\tilde{R}}{7\,\mu\, R_T^2\, R^4}}_{=:\, \beta(\xb)}\, G_0 + \frac{1}{\mu\, R^6}\left(\frac{18}{35}\,G_0^2 + \frac{27}{154}\,\Delta G^2\right)
\end{align}

\noindent
This is a quadratic function of $U = (G_0, \Delta G)^T$. Since the quadratic form $\mathcal{L}(\xb, U)$ is strictly convex in $U$ (Proposition~\ref{prop:stage_pos_def}(c)), the function $\Phi(\xb, \cdot)$ has a unique global minimum for each $\xb \in \mathcal{D}$.
\begin{align}
    \pd{\Phi}{G_0} = -\beta(\xb) + \frac{36}{35\,\mu\, R^6}\, G_0 = 0 \implies G_0^* = \frac{35\,\mu\, R^6}{36}\,\beta(\xb), \label{G0_star}\\
    \pd{\Phi}{\Delta G} = \alpha(\xb) + \frac{27}{77\,\mu\, R^6}\, \Delta G = 0 \implies \Delta G^* = -\frac{77\,\mu\, R^6}{27}\,\alpha(\xb) \label{DG_star}
\end{align}

\medskip
\noindent Step 4: Evaluate $\Phi^*(\xb) := \min_{U}\, \Phi(\xb, U) = \Phi(\xb, U^*)$.

\noindent
 Substituting~(Eq.\eqref{G0_star}-\eqref{DG_star}) in~(Eq.~\eqref{eq:Phi}) and using the identity for a minimized quadratic $\min_u (au^2 + bu) = -b^2/(4a)$, we obtain (carrying out the algebra for each control separately, since the cross-terms vanish due to the diagonal matrix):

For the $G_0$ component:
\begin{align*}
    -\beta\, G_0^* + \frac{18}{35\,\mu R^6}\,(G_0^*)^2 = -\frac{35\mu R^6}{36}\,\beta^2 + \frac{35\mu R^6}{72}\,\beta^2 = -\frac{35\mu R^6}{72}\,\beta^2
\end{align*}

For the $\Delta G$ component:
\begin{align*}
    \alpha\,\Delta G^* + \frac{27}{154\,\mu R^6}\,(\Delta G^*)^2 = -\frac{77\mu R^6}{27}\,\alpha^2 + \frac{77\mu R^6}{54}\,\alpha^2 = -\frac{77\mu R^6}{54}\,\alpha^2
\end{align*}
Therefore:
\begin{equation}\label{eq:Phi_star}
    \Phi^*(\xb) = -\frac{35\,\mu\, R^6}{72}\,\beta(\xb)^2 - \frac{77\,\mu\, R^6}{54}\,\alpha(\xb)^2
\end{equation}

\noindent Step 5: Show $\Phi^*(\xb) \leq 0$ with equality only at $\xb_T$.

\noindent
From~(Eq.~\eqref{eq:Phi_star}), since $\mu > 0$ and $R > 0$, we have $\Phi^*(\xb) \leq 0$ for all $\xb \in \mathcal{D}$. Moreover, $\Phi^*(\xb) = 0$ if and only if $\alpha(\xb) = 0$ and $\beta(\xb) = 0$. From the definitions:
\begin{align}
    &\alpha(\xb) = \frac{9\mathcal{A}\,(X - X_T)}{35\,\mu\, X_T^2\, R^4} = 0 \iff X = X_T, \qquad \\&
    \beta(\xb) = \frac{12\mathcal{B}\,(R - R_T)}{7\,\mu\, R_T^2\, R^4} = 0 \iff R = R_T.
\end{align}
Hence $\Phi^*(\xb) = 0$ if and only if $\xb = \xb_T$.

\medskip
\noindent Step 6: Conclusion.

\noindent
For part (a), from Steps 2 and 5 we can choose $U = U^*$ such that $\dot{\mathcal{T}}(\xb, U^*) < 0$ for $\xb \neq \xb_T$ (since $\Phi^* < 0$ and $\mathcal{L} \geq 0$, we have $\dot{\mathcal{T}} \leq \Phi^* - \mathcal{L} \leq \Phi^* < 0$). Hence terminal cost $\mathcal{T}$ is a CLF.

\noindent
For part (b), we show $\Phi^*(\xb) = \min_{U}\{\dot{\mathcal{T}}(\xb,U) + \mathcal{L}(\xb,U)\} \leq 0$ on all $\mathcal{D} \setminus \{\xb_T\}$, which is stronger than the local requirement. In fact, terminal cost is a cost-compatible CLF on domain $\mathcal{D}$.
\end{proof}

\noindent
\textbf{Remark:}
CLF compatibility condition on $\mathcal{D}$ is a consequence of the special structure of the droplet transport dynamics where system is purely control-driven, and the viscous (stage) cost is a positive definite quadratic in the controls. In many applications, the CLF condition can only be verified locally (near a linearized equilibrium). However, this property here opens the door to global stability results.

\subsection{Explicit Quantification of the CLF Decrease Rate}\label{sec:decrease_rate}

The following proposition provides an explicit lower bound on the decrease rate of $\mathcal{T}$ under the CLF-compatible control, which will be essential for establishing stability.

\begin{proposition}[Quadratic Decrease Rate]\label{prop:decrease_rate}
There exist positive constants $c>0$ (depending on $\mathcal{A}$, $\mathcal{B}$, $X_T$, $R_T$, $\mu$, $R_{\min}$, $R_{\max}$) such that for all $\xb \in \mathcal{K}$
\begin{equation}\label{eq:decrease_rate}
    \Phi^*(\xb) \leq -c\,\norm{\tilde{\xb}}^2,\nonumber
\end{equation}
and therefore
$\label{eq:CLF_Lyap_decrease}
    \min_{U}\left\{\dot{\mathcal{T}}(\xb, U) + \mathcal{L}(\xb, U)\right\} \leq -c\,\norm{\tilde{\xb}}^2$

\end{proposition}

\begin{proof}
From~(Eq.~\eqref{eq:Phi_star}):
\[
    \Phi^*(\xb) = -\frac{35\mu R^6}{72}\left(\frac{12\mathcal{B}\tilde{R}}{7\mu R_T^2 R^4}\right)^2 - \frac{77\mu R^6}{54}\left(\frac{9\mathcal{A}\tilde{X}}{35\mu X_T^2 R^4}\right)^2
\]
\begin{align}
    \Phi^*(\xb) &= -\frac{10\mathcal{B}^2\,\tilde{R}^2}{7\,\mu\, R_T^4\, R^2} - \frac{231\,\mathcal{A}^2\,\tilde{X}^2}{2450\,\mu\, X_T^4\, R^2} \label{eq:Phi_star_expanded}
\end{align}
On the compact set $\mathcal{K}$, we have $R \leq R_{\max}$, so $R^{-2} \geq R_{\max}^{-2}$. Therefore:
\[
    \Phi^*(\xb) \leq -\frac{1}{R_{\max}^2}\left(\frac{231\,\mathcal{A}^2}{2450\,\mu\, X_T^4}\,\tilde{X}^2 + \frac{10\,\mathcal{B}^2}{7\,\mu\, R_T^4}\,\tilde{R}^2\right) \leq -c\,\norm{\tilde{\xb}}^2,
\]
where
\begin{equation}\label{eq:c1_def}
    c := \frac{1}{R_{\max}^2}\min\left(\frac{231\,\mathcal{A}^2}{2450\,\mu\, X_T^4},\; \frac{10\,\mathcal{B}^2}{7\,\mu\, R_T^4}\right) > 0.
\end{equation}
\end{proof}

\noindent
\textbf{Remark:} The quadratic decrease rate quantifies the rate at which the CLF-compatible control dissipates the terminal cost. The constant `$c$' (Eq.~\eqref{eq:c1_def}) depends explicitly on the physical parameters of the droplet system; the fluid viscosity $\mu$, the target states $(X_T, R_T)$, and the admissible range of droplet widths $[R_{\min}, R_{\max}]$. This means that for a given microchannel configuration and transport task, the quadratic rate of decrease can be computed a priori, providing a quantitative design tool for selecting the weighting coefficients in the terminal cost.

\subsection{The CLF-induced stabilizing feedback}\label{sec:CLF_feedback}

\begin{definition}[CLF Controller]\label{def:CLF_controller}
The CLF controller (also called the \emph{pointwise min-norm controller}) is the feedback law
\begin{equation}\label{eq:CLF_controller}
    U_V(\xb) := \argmin_{U \in \R^2} \left\{\dot{\mathcal{T}}(\xb, U) + \mathcal{L}(\xb, U)\right\} = U^*(\xb),
\end{equation}
given explicitly by~(Eq.~\eqref{G0_star}--\eqref{DG_star}).

\[
   G_0^* = \frac{35\,\mu\, R^6}{36}\,\beta(\xb), \nonumber\\ \quad
   \Delta G^* = -\frac{77\,\mu\, R^6}{27}\,\alpha(\xb)
   \nonumber
\]
\end{definition}

\begin{proposition}[Closed-loop stability Under CLF feedback]\label{prop:CLF_stability}
The CLF controller~\eqref{eq:CLF_controller} exponentially stabilizes $\xb_T$ within $\mathcal{K}$. Specifically, under the closed-loop dynamics $\dot{\xb} = \fb(\xb, U_V(\xb))$:
\begin{equation}\label{eq:CLF_decrease_closed}
    \dot{\mathcal{T}} = \nabla\mathcal{T}(\xb)\cdot\fb(\xb, U_V(\xb)) \leq \Phi^*(\xb) \leq -c\,\norm{\tilde{\xb}}^2 \leq -\frac{c}{\overline{p}}\,\mathcal{T}(\xb),
\end{equation}
implying exponential decrease of $\mathcal{T}$ at rate $c/\overline{p}$, where $c$ is defined in Proposition~\ref{prop:decrease_rate}.
\end{proposition}

\begin{proof}
By construction, $\Phi^*(\xb) = \dot{\mathcal{T}}(\xb, U_V) + \mathcal{L}(\xb, U_V) \leq 0$. Since $\mathcal{L} \geq 0$, we have $\dot{\mathcal{T}}(\xb, U_V) \leq \Phi^*(\xb) \leq -c\norm{\tilde{\xb}}^2$. Using the upper bound $\mathcal{T}(\tilde{\xb}) \leq \overline{p}\norm{\tilde{\xb}}^2$ from~(Eq.~\eqref{eq:quadratic_bounds}), we get $\norm{\tilde{\xb}}^2 \geq \mathcal{T}/\overline{p}$, hence $\dot{\mathcal{T}} \leq -(c/\overline{p})\mathcal{T}$. By the comparison lemma, $\mathcal{T}(\tilde{\xb}(t)) \leq \mathcal{T}(\tilde{\xb}(0))\,e^{-c t/\overline{p}}$, which establishes exponential convergence of $\mathcal{T}$ to zero, and hence $\tilde{\xb}(t) \to \bm{0}$.
\end{proof}

\hfill $\blacksquare$


\noindent
\textbf{Remark:} This proposition confirms that the droplet reaches and remains at the target state with an exponential rate and that is directly tunable through the terminal cost weights. Physically, the CLF feedback simultaneously steers the droplet position (via $\Delta G$) and regulates its width (via $G_0$), and the exponential stability ensures that perturbations in either state are corrected at a decaying rate. This is particularly relevant for microfluidic applications, where disturbances from channel imperfections or flow fluctuations demand robust recovery to the desired operating state. Notably, this result holds over the physical domain $\mathcal{D}$ rather than a small neighborhood. It is a distinction from typical local CLF analyses that arises directly from the driftless, control-affine structure of the droplet dynamics.

The ratio $\mathcal{A}/X_T^2$ to $\mathcal{B}/R_T^2$ in the weighting matrix $P$ (Eq.~\eqref{eq:terminal_error}) controls the relative correction rates for position versus width errors under the CLF feedback. Balanced weights ($\mathcal{A}/X_T^2 \approx \mathcal{B}/R_T^2$) yield the fastest guaranteed convergence, since the exponential decay rate $c/\bar{p}$ is limited by the smaller diagonal entry of $P$.

\begin{figure}
    \centering
    \includegraphics[width=0.5\linewidth]{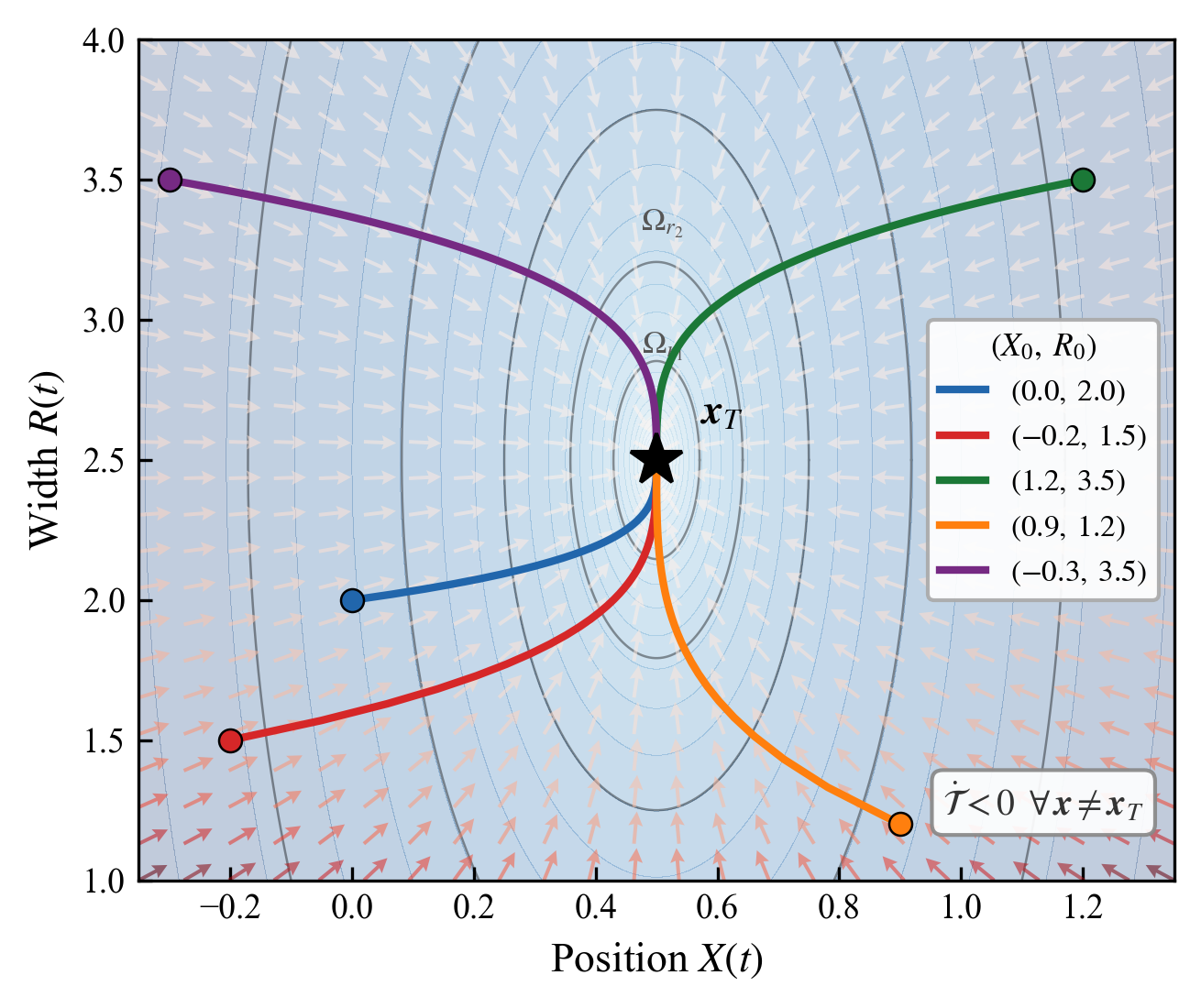}
    \caption{Closed-loop phase portrait under the CLF feedback~\eqref{eq:CLF_controller}. 
    Arrows show the vector field $\dot{\xb} = \fb(\xb, U_V(\xb))$. 
    Black ellipses denote sublevel sets $\Omega_r$ of $\mathcal{T}$. 
    Five trajectories from distinct initial conditions converge to $\xb_T$ (star), with the flow everywhere transversal to the level sets, confirming $\dot{\mathcal{T}} < 0$ on $\mathcal{D} \setminus \{\xb_T\}$. Parameters used: $\mu = 0.5$, $X_T = 0.5$, $R_T = 2.5$.}
    \label{fig:phase}
\end{figure}

\subsection{Numerical Verification}
To complement the stability structure established by Propositions~5.4 and~5.11, we simulate the closed-loop system $\dot{\xb} = \fb(\xb, U_V(\xb))$ under the CLF feedback (Eq.~\eqref{eq:CLF_controller}). Fig.~\ref{fig:phase} shows the resulting phase portrait for parameters $\mu = 0.5$, $X_T = 0.5$, $R_T = 2.5$, and $\mathcal{A} = \mathcal{B} = 10^3$. The vector field (arrows) is computed by evaluating the analytical CLF controller (Eqs.~\eqref{G0_star}--\eqref{DG_star}) and the closed-loop dynamics (Eq.~\eqref{eq:dynamics}) at each point in the state space. Five trajectories are integrated (using a Runge--Kutta solver) from initial conditions chosen to span all four quadrants of the error space $(\tilde{X}, \tilde{R})$, representing physically distinct scenarios: droplets that are too narrow, too wide, displaced left, or displaced right of the target.
 
There are two main features of the phase portrait: first, the vector field is everywhere transversal to the sublevel sets $\Omega_r$ of $\mathcal{T}$ (black ellipses) and points inward, providing geometric confirmation that $\dot{\mathcal{T}} < 0$ on $\mathcal{D} \setminus \{\xb_T\}$. Second, all five trajectories converge to $\xb_T$ regardless of their starting quadrant, consistent with the CLF property established above. The trajectories also reveal that the CLF controller simultaneously corrects position and width errors, with the relative correction rate governed by the ratio $\mathcal{A}/X_T^2$ to $\mathcal{B}/R_T^2$ in the terminal cost weighting matrix $P$ (Eq.~\eqref{eq:terminal_error}).
The controller itself is fully analytical (Eqs.~\eqref{G0_star}--\eqref{DG_star}); the numerical integration serves only to visualize the closed-loop behavior that these expressions produce.

\section{Reduced-order controller on the PDE system across
capillary regimes}
\label{si:gamma_sweep}

\begin{figure}[t]
    \centering
    \includegraphics[width=1\linewidth]{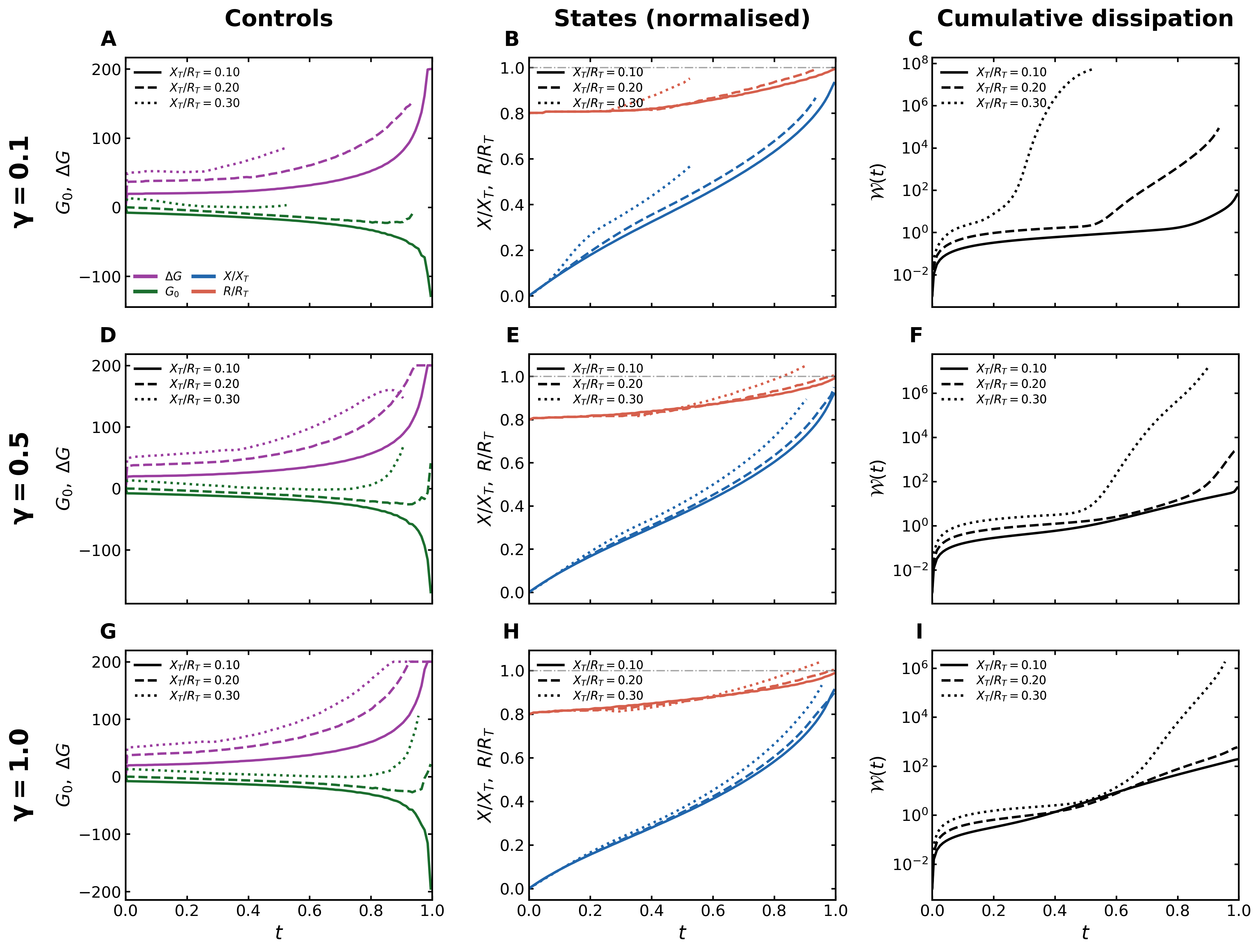}
    \caption{Reduced-order controller applied to the PDE system.
    Rows correspond to $\gamma = 0.1$, $0.5$, and $1.0$; columns to the
    controls $G_0(t)$ and $\Delta G(t)$, the states normalised by their
    targets, and the accumulated dissipation $\mathcal{W}(t)$.}
    \label{fig:figs3}
\end{figure}

The receding-horizon scheme of Section~V is repeated for
$\gamma = 0.1$, $0.5$, and $1.0$ at three transport ratios,
$X_T/R_T = 0.10$, $0.20$, and $0.30$. All other parameters are
unchanged including $R_0 = 2.0$, $R_T = 2.5$, $\mu = 0.5$,
$\mathcal{A} = \mathcal{B} = 10^3$, $T = 1$,
$\Delta t = 5\times10^{-3}$, and $N = 800$ elements on a domain of
length $L = 8$.
Fig~\ref{fig:figs3} resolves the mechanism behind the attainment
and cost trends. The antisymmetric control $\Delta G$ increases monotonically in every case. The reduced dynamics scale as $R^{-4}$, so the forcing needed to sustain translation grows as the droplet spreads. This increment is the slowest at $\gamma = 1.0$ where capillary restoring forces limit spreading, and steepest at $\gamma = 0.1$. 
Four of the nine simulations stop before the end of the horizon because the
film height loses positivity. At $X_T/R_T = 0.30$ this happens for all
three surface tensions, at $t \approx 0.54$, $0.91$, and $0.97$ for
$\gamma = 0.1$, $0.5$, and $1.0$ respectively, so stiffer interfaces
delay the failure without preventing it. At $X_T/R_T = 0.20$ only the
weakly capillary case breaks down, at $t \approx 0.94$, and the
remaining simulations complete the full horizon.

The accumulated dissipation $\mathcal{W}(t)$ is nearly flat over the
first half of the horizon and then rises by three to six orders of
magnitude. Transport cost is therefore concentrated in a short terminal interval, where the shrinking horizon drives the costate solution, and with it the controls, toward large values. 
The symmetric control $G_0$ is negative at $X_T/R_T = 0.10$ and
$0.20$, so the droplet stretches while it translates and follows the
translate--relax route, whereas at $X_T/R_T = 0.30$ it starts positive
for all three surface tensions and changes sign between
$t \approx 0.4$ and $0.7$. Both transport strategies of Section~III
are therefore recovered from the continuum state, with the switch
between them set by the target displacement.

\newpage

\bibliographystyle{IEEEtran}
\bibliography{citation}